\documentstyle[12pt,aasms4]{article}
\begin{document}
\baselineskip=24pt

\title{EVOLUTION OF X-RAY CLUSTERS OF GALAXIES AND SHOCK HEATING OF 
       INTRACLUSTER MEDIUM}
\author{{\sc Motokazu Takizawa and Shin Mineshige}}
\affil{Department of Astronomy, Faculty of Science, Kyoto University, 
       Sakyo-ku, Kyoto 606-01,JAPAN}
\authoremail{takizawa@kusastro.kyoto-u.ac.jp}

\begin{abstract}
Evolutions of X-ray clusters of galaxies are studied by 
N-body (shell model) + mesh code (TVD) simulations on the
assumption of spherical symmetry.
We consider a density perturbation of 
$10^{15} M_{\odot}$ composed of dark matter and gas
%These perturbations correspond to $\sim 1.5 \sigma$ 
in cold dark matter dominated universe with the cosmological density 
parameter, $\Omega_0 = 1$ or $0.2$.
%where $\sigma$ is rms density fluctuation on a cluster scale.
A shock front appears during its initial collapse,
moving outward as ambient gas accretes towards cluster center.
The shock front separates the inner X-ray emitting, hot region, 
where gas is almost in hydrostatic equilibrium but with 
small radial infall ($\sim 100$km s$^{-1}$) being left,
from the outer cool region, where gas falls
almost freely and emits no X-rays. 
Gas inside the shock is strongly compressed and heated by
shock so that X-ray luminosity rapidly rises in the early stage
(until temperature reaches about virial).
In the late stage, however, 
the X-ray luminosity rises only gradually
due partly to the expansion of
the inner high temperature region and partly to
the increase of X-ray emissivity of gas as the results of
continuous adiabatic compression inside the shock.
We also find for clusters in lower density universe that
the density distribution is generally less concentrated 
and, hence, X-ray luminosity more slowly rises than in higher 
density universe.

The shock front structure, which was not clearly resolved
in the previous SPH simulations, is clearly captured 
by the present simulations.  Our results confirm 
%that hot gas of X-ray cluster 
%of galaxies is separated with a definite boundary and 
that shock heating plays 
an important role in the heating process of intracluster medium.
In addition, we find a sound wave propagating outward, 
thereby producing spatial
modulations with amplitudes of $\sim$ 10 \% 
in the radial temperature and density profiles and
time variations in the strength of the shock.
Such modulations, if observed, could be used as a probe to 
investigate the structure of clusters.
\end{abstract}


\section{INTRODUCTION}

Clusters of galaxies (CGs) are luminous X-ray sources (Sarazin 1988).
X-rays are 
thought to be emitted through thermal bremsstrahlung from very hot,
optically thin plasma gas in intergalactic space, 
which is called intracluster medium (ICM).
Recent X-ray and optical observations have revealed dynamical aspects 
of CGs (Fabricant, Kent \& Kurtz 1989; Kneib et al. 1995).  
Time evolution of the cluster X-ray luminosity function has been 
confirmed; each CG became brighter and the number density 
of CGs with a fixed X-ray luminosity
increased through mergers towards zero redshift
% by the observations of distant clusters 
(Edge et al. 1990; Gioia et al.1990; Henry et al. 1992; Bower et al. 
1993; Castander et al. 1994).  
To understand the physical structure of CGs, therefore,
it is essential
%necessary not only to study equilibrium structure of CGs but also 
to consider non-equilibrium processes involved with
the dynamical evolution of CGs.

Recent extensive X-ray observations have established 
that the temperature profiles of ICM 
are nearly flat in many clusters (Fabian 1994; Ohashi et al 1996). 
Although there is no widely accepted explanation for this,
%as to why ICM is nearly isothermal, 
it is, at least, reasonable to expect that 
shock heating through gravitational collapse plays an important role.
It then follows that the isothermality of gas may reflect 
the shape and time evolution of the gravitational potential well.
%the DM density distribution which determines a gravitational potential.
What then governs the total dynamics of CGs?
It is galaxies and dark matter (DM),
both of which can be regarded
as a collisionless self-gravitating many-body system.
Dynamics of ICM, collisional fluid on a CG scale, simply follows
the dynamical evolution of the gravitational potential exerted
mainly by DM and galaxies.
Thus, the thermal history of ICM is rather sensitive to the physical
processes involving violent time variation of gravitational potential 
field (such as formation, merger, etc), 
%since then  shock heating is expected to occur. 
Violent relaxation 
(Lynden-Bell 1967) is thought to play a crucial role there,
%for evolution of collisionless self-gravitating systems 
although its physical significance is still poorly understood 
(Funato, Makino \& Ebisuzaki 1992a,b; Takizawa \& Inagaki 1996).

CGs were believed to be formed from overdense perturbations 
in the universe.
They grew through gravitational instability and collapsed at
$z \sim 1$.  CGs have been growing by accreting ambient matter, 
which is still an ongoing process at the present. 
Gunn \& Gott (1972) quantitatively
discussed the growth of a spherical symmetric perturbation consisting 
of only collisionless particle in an expanding universe.
Bertschinger (1985) found the self-similar solution 
describing evolution of spherical density fluctuation
consisting of DM and gas
in the Einstein-de Sitter universe (where the cosmological density
parameter is $\Omega_0=1$ and the cosmological constant is $\lambda_0=0$).

There have been plenty of
numerical studies performed so far regarding the 
formation and evolution of CGs by using N-body and hydrodynamic codes.  
Perrenod (1978) was the first to calculate
the evolution of a spherical symmetric cluster
with standard mesh hydrodynamic code.  Three-dimensional 
calculations have been carried out recently, which mostly use
smooth particle hydrodynamics (SPH) codes
(Evrard 1990; Thomas \& Couchman 1992; Bryan et al. 1994; 
Metzler \& Evrard 1994; Navarro, Frenk \& White 1995, hereafter NFW).
 According to Evrard (1990), a shock front moved outward when gas around
the cluster accreted onto the cluster center and 
a relatively flat temperature profile was realized within the shock front. 
This result was significant in the sense that
the above expectation was confirmed qualitatively.  
However, quantitative estimation may be problematic because of
limited spatial resolutions
and poor reliability of SPH code in calculations of shocks.
It might be kept in mind that
SPH codes are easy to work with and could give reasonable accuracy,
however, they are not better for problems with discontinuities 
(such as strong shock and contact discontinuity) 
than mesh codes (Monaghan 1992).
On a larger scale, numerical simulations of cosmological structure formation
by using N-body and hydrodynamic mesh code have been carried out mainly to 
investigate the statistical properties of CGs
(Cen \& Ostriker 1994; Kang et al. 1994; Anninos \& Norman 1996).

In this paper, we focus on the dynamical aspects of ICM,
such as time-dependent properties of shock waves
and the effects of shock heating on the evolution of CGs.
For this purpose, we perform numerical simulations of spherical
CGs with N-body (shell model) + mesh code (TVD).
Note that the TVD code is one of the most useful tools to 
deal with spatial discontinuities.
Since we assume spherical symmetry,
the problem can be reduced to one-dimensional.
Better spatial resolutions can be achieved, therefore.
Using these codes, we calculate
dynamical evolutions of a density perturbation of $10^{15} M_{\odot}$
consisting of DM and gas and collapsing at $z \sim 1$ 
in universe with $\Omega_0 = 1$ or $0.2$.
It is also interesting to investigate
how different cosmological models affect evolution of CGs.

The rest of this paper is organized as follows. In section 
\ref{simulations} we describe our numerical methods and 
the adopted initial conditions.
In section \ref{results} we present our results, 
discussing physical processes underlying
the structural evolution of calculated CGs. 
In section \ref{summary} we summarize our results
and discuss their implications.

\section{THE SIMULATIONS}\label{simulations}

In the present study we regard CG consisting of two components:
collisionless particles corresponding to galaxies and DM, and
gas corresponding to ICM.
When calculating gravity both components are considered,
although the former component dominates over the latter.
We also assume spherical symmetry in all the calculations.

\subsection{Basic Equations for Collisionless Particles}

For calculations of collisionless particles
we adopt a shell model (H\'enon 1964).
The distribution function of spherical symmetric stellar systems can be 
expressed as $f(r,u,v,t)$ , where $r$ is radial distance, $u$ is radial
velocity, $v$ is tangential velocity, and $t$ is time, respectively. 
So the state of the system is represented by $N$ points in the $(r,u,v)$
space (with $N$ being the number of shells) and
a trajectory of the $i$-th point at $(r_i,u_i,v_i)$ is 
calculated according to the equations of motion;
 \begin{eqnarray}
  \frac{dr_i}{dt} &=& u_i , \label{eq:drdt}  \\
  \frac{du_i}{dt} &=& \frac{A_i^2}{r_i^3} - \frac{G M_i}{r_i^2}  , \label{eq:dudt}
 \end{eqnarray}
for $i=1 \sim N$,
where $G$ is the gravitational constant, $A_i \equiv r_i v_i$ 
(= constant in time) is angular momentum of the $i$-th shell,
and $M_i$ is total mass (also including mass of gas) interior to $r_i$, 
respectively.
Since it is convenient to carry out numerical calculations 
using comoving coordinates for our purpose, we transform $(r_i,u_i,v_i)$ to 
$(R_i,U_i,v_i)$ as follows;
 \begin{eqnarray}
 r_{i} &=& a(t) R_{i} ,  \label{eq:rvsR} \\
 u_{i} &=& \dot{a} R_{i} + U_{i} ,  \label{eq:uvsU}
 \end{eqnarray}
where $a(t)$ is the dimensionless scale factor of the universe
and $\dot{a}$ represents the derivative of $a(t)$ with respect to time.
Equations (\ref{eq:drdt}) and (\ref{eq:dudt}) are then transformed into
 \begin{eqnarray}
 \frac{d R_{i}}{d t} &=& \frac{U_{i}}{a}, \label{eq:dRdt}  \\
 \frac{d U_{i}}{d t} &=& \frac{A_{i}^2}{(a R_{i})^3} 
                       - \frac{G M_{i}}{(a R_{i})^2} 
                       - \frac{\dot{a}}{a} U_{i} - \ddot{a} R_{i},
                                                               \label{eq:dUdt}
 \end{eqnarray}
respectively, by using equations (\ref{eq:rvsR}) and (\ref{eq:uvsU}).
Equations (\ref{eq:dRdt}) and (\ref{eq:dUdt}) are integrated by
using leap-frog method. 
As to the inner boundary condition
we set a reflecting wall at $r_{\rm rw}=0.02/(1+z)$ and impose that
when a shell reaches the wall, $r_i < r_{\rm rw}$,
that shell is elastically reflected.
Only shells with rather small angular momentum is influenced by the wall.

\subsection{Basic Equations for Gas}

For gas components basic equations in the comoving frame are as follows,
 \begin{eqnarray}
 \frac{\partial \rho}{\partial t} 
        + \frac{1}{a} \frac{\partial}{\partial R} ( \rho v_{\rm gas} ) 
    &=& -3 \frac{\dot{a}}{a} \rho - \frac{2}{a R} \rho v_{\rm gas}, \\
 \frac{\partial}{\partial t} (\rho v_{\rm gas}) 
        + \frac{1}{a} \frac{\partial}{\partial R} (\rho v_{\rm gas}^2 + P) 
    &=& -4 \frac{\dot{a}}{a} \rho v_{\rm gas} - 2 \frac{\rho v_{\rm gas}^2}{a R} 
        + \rho g_R, \\
 \frac{\partial}{\partial t} (\rho E) 
        + \frac{1}{a} \frac{\partial}{\partial R} (\rho H v_{\rm gas}) 
    &=& - \frac{\dot{a}}{a} \biggr( \frac{5}{2} \rho v_{\rm gas}^2 
                                   - \frac{3 \gamma}{\gamma - 1} P \biggl) 
        - 2 \frac{\rho H v_{\rm gas}}{a R} + \rho g_R v_{\rm gas},
 \end{eqnarray}
where $\rho$ is gas density, $v_{\rm gas}$ is radial velocity of gas,
$P$ is gas pressure, and $\gamma$ is the adiabatic exponent, respectively.
The total energy of gas per unit mass, $E$, and the enthalpy 
per unit mass, $H$, are given by
 \begin{eqnarray}
 E &=& \frac{P}{\rho (\gamma - 1)}+\frac{v_{\rm gas}^2}{2} ,  \\
 H &=& \frac{\gamma}{\gamma - 1} \frac{P}{\rho} + \frac{v_{\rm gas}^2}{2},
 \end{eqnarray}
respectively, and $g_R$ is defined by
 \begin{eqnarray}
 g_R \equiv - \frac{G M_R}{(a R)^2} - \ddot{a} R,
 \end{eqnarray}
where, $M_R$ is the total mass (including gas and collisionless particles)
inside $R$.

We neglect viscosity and angular momentum of gas. We also assume that 
gas is ideal gas with  $\gamma = 5/3$.
A second order up-wind TVD code (minmod limiter)
is used for our simulations (Hirsch 1990).  Number of mesh point is 500 
and one mesh spacing corresponds to $\triangle r = 0.02/(1+z)$ Mpc
with $z$ being a redshift [$1+z = a(0)/a(t)$].  
We set boundary condition as follows. 
The inner edge is assumed to be a perfectly reflecting point;
 \begin{eqnarray}
  \rho_{-1} = \rho_{0} = \rho_{1},   \\
  P_{-1} = P_{0} = P_{1},   \\
  v_{{\rm gas},-1} = v_{{\rm gas},0} = 0 ,
 \end{eqnarray}
where $\rho_i$, $P_i$ and $v_{{\rm gas},i}$ is gas density, gas pressure
and radial velocity of gas at the $i$-th mesh point, and
the first mesh point corresponds to the inner boundary.
The $0$-th and $-1$-st points are necessary for calculations
(to derive spatial derivatives of physical quantities).
The outer edge is assumed to be a perfectly transmitting surface;
 \begin{eqnarray}
  q_{n+2} &=& 3 q_{n} - 2 q_{n-1},  \\
  q_{n+1} &=& 2 q_{n} -  q_{n-1},  
 \end{eqnarray}
where $q_i$ is any physical quantity of gas at the $i$-th mesh point
and the $n$-th corresponds to the outer boundary.
Again, the $(n+1)$-th and $(n+2)$-th points are necessary 
for calculation purpose.

\subsection{Models and Initial Conditions}\label{models}

In this paper, calculations are carried out from $z_{\rm ini}=10$
to the present time ($z_0=0$).
We adopt cosmological models with no cosmological constant, 
$\lambda_0 = 0$.  When $\Omega_0 = 1$, therefore, we find
 \begin{eqnarray}
 a(t) = \biggr(\frac{t}{\frac{2}{3} H_0^{-1}} \biggl)^{\frac{2}{3}},
 \end{eqnarray}
while when $0 < \Omega_0 < 1$ we have
 \begin{eqnarray}
 a(t) &=& \frac{ \Omega_0 }{ 2 ( 1 - \Omega_0 ) } ( \cosh \psi -1 ), \\
 H_0 t &=& \frac{ \Omega_0 }{ 2 ( 1 - \Omega_0 )^{3/2} } ( \sinh \psi - \psi ),
 \end{eqnarray}
where $H_0$ is the Hubble constant and is set to be
$H_0 = 100$ (km s$^{-1}$ Mpc$^{-1}$) in transformations of length and time.

We make initial density profiles in the same manner as Peebles (1982). 
At first we prepare $N$ concentric shells with uniform density being equal
to the mean density of the universe at $z_{\rm ini}=10$. 
Then a density fluctuation is introduced 
by perturbing the radius and velocity of each shell as
 \begin{eqnarray}
  r_i &=& r_i^{(0)} \left[ 1 + \delta ( r_i^{(0)} ) \right] ,
 \end{eqnarray}
where $r_i^{(0)}$ is the unperturbed coordinate and
$\delta(r)$ represents the perturbation as a function of $r$ (specified below).
The velocity perturbation is
written, using Zel'dovich approximation, by
 \begin{eqnarray}
  u_i &=& H(z_{\rm ini}) r_i^{(0)} \left[ 1 + 2 \delta (r_i^{(0)})\right] ,
 \end{eqnarray}
where $H(z_{\rm ini})$ is the Hubble constant at the initial epoch, and
we used $\delta \propto t^{2/3}$;
the relation which holds exactly in the Einstein-de Sitter universe.
The functional form of the perturbation is assumed as 
 \begin{equation}
  \delta ( r ) = \left\{
                \begin{array}{@{\,}ll}
                       - \frac{ \delta_0 }{3} \cos^2 \bigl( \frac{ \pi }{2} \frac{r}{r_0} \bigr) & \mbox{($0 \le r \le r_0$)} \\
                       0  & \mbox{($r_0 <  r$)}
                       \end{array}
	                     \right. \label{eq:nt} ,
 \end{equation}
where $r_0$ and $\delta_0$ are, respectively, 
the initial size of the fluctuation on the comoving scale 
and the parameter for displacement (see Nakamura 1996).
We take $(\delta_0, r_0) = (0.4, 9.0$ Mpc) 
for Model A (with $\Omega_0 = 1.0$) and
$(\delta_0, r_0) = (0.5, 15.4$ Mpc) 
for Model B (with $\Omega_0 = 0.2$), respectively.  In both models
the fluctuation contains a mass of about $10^{15} M_{\odot}$ and
correspond to typical density peaks with $\sim 1.5 \sigma$
in the CDM power spectrum with $\sigma_8 = 0.96$.  Here,
$\sigma$ is the rms density fluctuations on a cluster scale and
$\sigma_8$ is the rms of mass fluctuation 
%in randomly placed spheres with a radius of
on scale 8 Mpc; $\sigma_8 = 0.96$ was obtained 
from the observation of nearby galaxy distribution (see Suto 1993).
The profile of the initial density perturbation and the ratio of
$(\delta M / M)/\sigma_{8,{\rm CDM}}$ 
are illustrated in figure \ref{indmden} for Model A
(by the solid line) and B (by the dotted line), respectively.  Here,
$(\delta M / M)$ is the mass fluctuation averaged over 
a scale $r$ ,whereas $\sigma_{8,{\rm CDM}}$ is the
rms of mass fluctuation averaged over a scale $r$ 
obtained from CDM power spectrum with $\sigma_8 = 0.96$.

\begin{figure}[htbp]
 \caption[]{}
 \label{indmden}
\end{figure}% 

The initial conditions of gas are set as follows. 
At first density of gas was everywhere set to be 
1/10 of the mean density of the universe at $z=10$ and
temperature of gas ($T_{\rm gas,i}$) was everywhere taken to be $10^7$ K
except in Model LT, where we assumed $T_{\rm gas,i} = 10^6$K initially.
Then adiabatic fluctuation (cf. Eq. 16 and 17) was
imposed so that gas density became always 1/10 of that of DM.
Note that gas temperature distribution becomes nonuniform, accordingly.
We calculated 3 models in total. Model parameters are listed in
table \ref{parameters}.

\section{RESULTS}\label{results}

\subsection{Overview of Evolution}

At first we overview the evolution of Model A cluster.
The evolution of density and radial velocity profiles of DM are
displayed in figure \ref{dms8o1.0ex}. The different types of
lines correspond to different redshifts 
(and different times): $z=2.5$ ($t=1$ Gyr) by the long dash line;
$z=1.2$ ($t=2$ Gyr) by the short dash line; 
$z=0.4$ ($t=4$ Gyr) by the dotted line; and 
$z=0$ ($t=6.5$ Gyr) by the solid line, respectively.

\begin{figure}[htbp]
  \caption[]{}
  \label{dms8o1.0ex}
\end{figure}% 

Density profile basically obeys a power law, 
$\rho_{\rm DM} \propto r^{-2.6}$,
and evolves in a self-similar fashion after $z \simeq 1$.
The evolutions in the radial distributions of
radial velocity, density, temperature, pressure, 
and entropy (each of the gas component) are summarized in figure \ref{gascla1}.
The different types of lines 
correspond to same redshifts as in figure \ref{dms8o1.0ex}.
\begin{figure}[htbp]
  \caption[]{}
  \label{gascla1}
\end{figure}% 

There are fundamental features commonly seen in the course of evolution
of all the models, which can be summarized as follows.

 \begin{enumerate}

 \item Before the initiation of DM collapse (at $z \geq 2.5$; $t \leq 1$ Gyr) 
       gas continues to expand, following the cosmological expansion.  
       When DM begins to collapse, a shock wave emerges in the central part 
       and moves outwards, accreting ambient gas towards the center.
 \item The shock front separates the inner hot region from the outer
       cool region.
       In the inner region, gas is almost in hydrostatic equilibrium,
       although bulk velocity of radial infall$\sim 100$ km s$^{-1}$ 
       still remains. 
       The temperature profile is relatively flat and gas 
       is hot enough to emit X-ray.
       In the outer region, in contrast,
       gas falls almost freely and is too cool to emit X-ray. 
 \item Density profile of gas evolves self-similarly as 
       $n_{\rm gas} \propto r^{-2.4}$ inside the shock front except 
       near the center, where density profile is rather flat.
       Even after the passage of the shock, density, temperature,
       and, therefore, pressure, gradually increase with time.
       This indicates that the inner region is not perfectly in 
       hydrostatic equilibrium (will be discussed in \S 3.3).
 \item Entropy profile shows an overall increase outwards, suggesting
       larger entropy production taking place as the shock moves outward.
       There are wavy features seen, especially, in the distributions of 
       temperature and entropy.  These seem to be related to sound wave
       propagation (will be discussed in \S 3.4).
 \end{enumerate}

Let us next examine each item in more details and
discuss similarities and differences between different models.

\subsection{Density Profiles}

Density profiles of DM at $z=0$ can well be fitted with the $\beta$-model;
 \begin{eqnarray}
 \rho_{\rm DM} (r) = \rho_{{\rm DM},0} \biggr[1 + \Bigr( \frac{r}{r_{\rm DM}} \Bigl)^2 \biggl]^{-3 \beta_{\rm DM} /2}.
                                                               \label{eq:beta1}
 \end{eqnarray}
Here, $r_{\rm DM}$ and $\beta_{\rm DM}$ are fitting parameters.
We fit the results of simulations inside $r_{100}$, the
radius where the mean interior density is $100$ times of critical density
at $z=0$. The results of the fitting are summarized in table \ref{dmden}.
Note that these results may depend on initial density profile.
 
In the same way density profiles of gas at $z=0$ inside the shock fronts
can be fitted with the $\beta$-model,
 \begin{eqnarray}
 n_{\rm gas} (r) = n_{{\rm gas},0} \biggr[1 + \Bigr( \frac{r}{r_{\rm gas}}\Bigl)^2 \biggl]^{-3 \beta_{\rm gas} /2},
                                                                \label{eq:beta2}
 \end{eqnarray} 
where $r_{\rm gas}$ and $\beta_{\rm gas}$ are fitting parameters.
The results are listed in table \ref{gasden}.
In all models we find $\beta_{\rm DM} \approx \beta_{\rm gas} \sim 0.9$
and $r_{\rm DM} \approx r_{\rm gas}$.

We also depict nondimensional density profiles at $z=0$ in figure \ref{ndden} 
for Model A (by the solid line) and Model B (by the dotted line), respectively,
where density is scaled with $\rho_{{\rm c}0}$, critical density of 
the universe at $z=0$, and radius is scaled with $r_{100}$.
Nondimensional DM density profiles look similar among two models, whereas
gas component expands slightly in Model B, compared with that in Model A.
%This is because the total overdensity is taken to be same for both models
%although mean density ($\rho_0$) is smaller in Model B.

\begin{figure}[htbp]
 \caption[]{}
 \label{ndden}
\end{figure}% 

Let us finally compare our results with the previous results (NFW).
As for the density profiles of the clusters in Einstein-de Sitter Universe
(Models A and LT), the core radii are smaller than those of
NFW (who obtained $r_{\rm DM} \sim r_{\rm gas} \sim 0.2$),
while $\beta_{\rm DM}$ and $\beta_{\rm gas}$ are similar.
%These tendencies  are commonly seen in both DM and gas. 
On the other hand,
%central densities of DM and gas are different; 
central DM density, $\rho_{{\rm DM},0}$, is similar to that of NFW,
whereas central gas density, $\rho_{{\rm gas},0}$, of model A is
lower than that of NFW (our model LT is consistent with their model;
$n_{{\rm gas},0} \sim 5 \times 10^{-2}$).
This can be explained in terms of different initial gas temperatures
(see section \ref{ditg}).

\subsection{Temperature Profiles} 

Temperature and entropy profiles are shown in figure
\ref{ent} in the upper and lower panels, respectively.
The different types of lines correspond to different models: Model A
by the solid line; Model B by the dotted line; and Model LT by the short
dash line, respectively.  There are common characters as follows.
Temperatures gradually fall outwards until the shock front.  
There are small temperature fluctuations seen.  If we evaluate
the propagation speed of fluctuation pattern, it is 
$\sim 700 {\rm km}/{\rm s}$, of the order of the sound velocity 
inside the shock.  In addition, entropy pattern does not change much
during the course of wave propagation, indicating that structural
variation is adiabatic.
We may thus conclude that the temperature fluctuation
arises due to sound wave propagation.  

Inside the shock fronts temperature is nearly virial,
high enough to emit X-ray.  Entropy monotonically rises from the
center to the shock front. That is, near the center gas is heated up
mainly through adiabatic compression, thus possessing relatively lower
entropy.  Gas in the outer region, on the other hand, will eventually
be heated through shocks, which effectively transform kinetic energy of
accreting gas into thermal energy at their surface.  This explains
relatively large entropy just inside the shock.  
Note that regions with entropy decreasing outward are convectively
unstable. Convective motions, if occur, will smear out such a feature.

\begin{figure}[htbp]
 \caption[]{} \label{ent}
\end{figure}% 

To understand why temperature, density and entropy steadily increase 
with time even after the passage of the shock front 
and how sound waves arise and propagate outward, 
we check the balance between pressure gradient and gravity
in the inner parts.
Figure \ref{tsuriai} shows the ratio of the absolute value of
pressure gradient force, $|dP/dr|$, and
gravitational force inside the shock surface,
$F_{\rm g} \equiv G M_{r} \rho_{\rm gas} / r^2$,
from $t=5.5$ Gyr to $6.5$ Gyr.
If the system is perfectly 
in hydrostatic equilibrium, $|dP/dr|/F_{\rm g}$ should be unity.

This figure shows a clear tendency that
gravitational force overcomes pressure gradient force 
inside the shock front, thus inducing radial gas inflow from outside.
Therefore, gas is adiabatically compressed by the infalling material 
from outside.  
Importantly, the ratio changes with time. 
Since DM density profile hardly changes with time near the core, 
so does the gravity force;
the ratio changes are purely due to time and spatial variations
in pressure profiles.  When ambient gas suddenly falls towards the
center, pressure at the core abruptly increases.  This gives rise
to an outwardly propagating sound wave, since the central point
is a reflecting boundary due to spherial symmetry (cf. \S 2.2).
Figure \ref{wave} shows radial velocity profiles of gas inside the shock
from $t=5.5$ Gyr to $6.5$ Gyr.
We confirm that radial infall systematically remains even after 
the passage of shock.  In addition we confirm fluctuation pattern 
propagating outwards with a speed roughly equal to the sound velocity.

\begin{figure}[htbp]
  \caption[]{}
  \label{tsuriai}
\end{figure}% 

\begin{figure}[htbp]
  \caption[]{}
  \label{wave}
\end{figure}%

In our simulations thermal conduction is neglected.  It is possible,
however,  that thermal conduction, if efficient,
will erase temperature fluctuations on a small scale.  
It is thus worthwhile evaluating conduction
timescale in the simulated clusters.  
Conduction time scale is generally expressed as (cf. Sarazin 1988)
%$t_{{\rm cond}}$ is,
\begin{eqnarray}
 t_{{\rm cond}} \approx \frac{n_{\rm e} l_{\rm T}^2 k_{\rm B}}{\kappa},
 \label{eq:tcond}
\end{eqnarray}
where $n_{\rm e}$ is electron number density, $l_{\rm T}$ is scale 
length of temperature gradient, $k_{\rm B}$ is the Boltzmann constant, 
and thermal conductivity for hydrogen plasma is (Spitzer 1962)
\begin{eqnarray}
 \kappa \approx 4.6 \times 10^{13} \biggr( \frac{T_{\rm e}}{10^8 {\rm K}}
\biggl)^{5/2} \biggr( \frac{\ln \Lambda }{40} \biggl)^{-1} ({\rm erg
s}^{-1} {\rm cm}^{-1} {\rm K}^{-1}), \label{eq:kappa}
\end{eqnarray}
where $\ln \Lambda$, Coulomb logarithm, is
\begin{eqnarray}
 \ln \Lambda = 37.8 + \ln \biggr[ \Bigr( \frac{T_{\rm e}}{10^8 {\rm K}}
\Bigl) \Bigr( \frac{n_{\rm e}}{10^{-3} {\rm cm}^{-3}} \Bigl)^{-1/2} \biggl]. 
\label{eq:coulomb}
\end{eqnarray}
From equations (\ref{eq:tcond}) and (\ref{eq:kappa}), we derive
%$t_{{\rm cond}}$ is numerically,
\begin{eqnarray}
 t_{{\rm cond}} \approx 3 \times 10^7 {\rm yr} \biggr( \frac{n_{\rm e}}{10^{-3}
{\rm cm}^{-3}} \biggl) \biggr( \frac{T_{\rm e}}{10^8 {\rm K}} \biggl)^{-5/2}
\biggr( \frac{l_{\rm T}}{0.1 {\rm Mpc}} \biggl)^2 \biggr( \frac{\ln
\Lambda}{40} \biggl). \label{eq:ntcond}
\end{eqnarray}

For Model A, for example, we find 
$t_{{\rm cond}} \sim 1.5 \times 10^{8} {\rm yr}$
at $r\approx 0.1$ Mpc (where $n_{\rm e} \sim 10^{-3} {\rm cm}^{-3}$, 
$T_{\rm e} \sim 5 \times 10^7$ K, and $l_{\rm T} \sim 0.1$ Mpc)
and $t_{{\rm cond}} \sim 1.5 \times 10^7 {\rm yr}$ 
at $r \approx 1 {\rm Mpc}$ 
(where $n_{\rm e} \sim 10^{-5} {\rm cm}^{-3}$, 
$T_{\rm e} \sim 2 \times 10^7$ K, and $l_{\rm T} \sim 0.1$ Mpc).
At both radii, hence, conduction seems to be efficient.
The same is true for Model B.  However, we should note that
the usage of the classical conductivity (Spitzer 1962)
is in question for CGs.  In fact, we cannot explain 
the existence of cooling flows as long as we employ the classical one
(Binney \& Cowie 1981).  Instead,
it is suggested that tangled magnetic field
(Rosner \& Tucker 1989) or plasma instabilities 
(Pistinner \& Shaviv 1996) are likely to suppress heat conduction 
significantly in CGs.  Temperature fluctuations can then survive.

\subsection{Evolution of Shock Surface}

As we have seen in figure \ref{wave}, sound wave propagates outward,
modulating temperature profile. 
This also affects shock front propagation, since sound velocity 
inside the shock is three times greater than the front velocity
in strong shock limits.
We illustrate in figure \ref{shock} the radius of a shock surface 
($r_{\rm shock}$) as a function of the
look-back time, time measured from the present time;
i.e., $t_{{\rm lb}}=0$ corresponds to the present time ($z=0$).
The different types of lines correspond to same
models as in figure \ref{ent}.  Shock surface moves with
nearly a constant velocity ($\sim 200$km s$^{-1}$).  The radius of
shock surface at $z=0$, $r_a$, and mean propagation velocity, $v_a$,
are listed in table \ref{shockpara}.  Again, there are wavy features
seen in this figure.
\begin{figure}[htbp]
 \caption[]{} \label{shock}
\end{figure}% 

To understand the physics causing modulating features, we plot time
variation of shock strength and shock radius ($r_{{\rm shock}}$) for Model A 
in figure \ref{shockamp}.  To evaluate shock strength we use
$\Delta v = v_{{\rm in}} - v_{{\rm out}}$
with $v_{{\rm in}}$ and $v_{{\rm out}}$ being
radial gas velocities inside and outside the shock surface, respectively,
in the upper panel, %in figure \ref{shockamp} (a) 
and $T_7 \Delta S$ where $T_7$ is the
pre-shock gas temperature in the unit of $10^7 {\rm K}$ and 
$\Delta S = S_{{\rm in}} - S_{{\rm out}}$ with $S_{{\rm in}}$ and
$S_{{\rm out}}$ being specific entropies inside and outside
the shock surface, respectively, in the lower panel. 
Since this quantity is proportional to heat produced through shock heating, 
it is a good representation of the shock strength.
Both panels show time modulation in shock strengths.
This in turn creates spatial modulation in entropy profile,
since shock radius moves outward with time; namely,
at the radius over which the shock passed with
its maximum (minimum) strength in time,
radial entropy profile exhibits a rapid (or slow) rise outward.
The time modulation in shock strengths is likely to be caused by
shock radius oscillation, and this oscillation is, 
as we discussed above, caused by the sound wave propagation.
Note that while the wavy pattern in temperature profile at a fixed 
radius varies with time because of sound wave propagation,
that in entropy profile does not,
since entropy profile is unaffected by sound waves.

\begin{figure}[htbp]
 \caption[]{} \label{shockamp}
\end{figure}%

\subsection{Evolution of X-Ray Luminosity}

Time evolutions of X-ray luminosity, $L(t_{{\rm lb}})$, and
the normalized luminosity,
$L_{\rm n}(t_{\rm lb}) \equiv L(t_{\rm lb})/L(t_{\rm lb}=0)$,
are plotted for each model in figure \ref{tl} (a) and (b), 
respectively. When calculating luminosity,
we assume thermal bremsstrahlung of optically thin plasma (Rybicki \&
Lightman 1979), 
\begin{eqnarray} 
 \varepsilon^{\rm ff} \equiv
 \frac{dW}{dV\,dt} = 1.4 \times 10^{-27} T^{1/2} n^2 \bar{g}_B 
                          ~({\rm erg~s}^{-1}{\rm cm}^{-3}), 
\label{eq:eff} 
\end{eqnarray} 
where $T$ is
gas temperature, $n$ is number density, $\bar{g}_B$ is a frequency
average of velocity-averaged Gaunt factor. In this paper we set
$\bar{g}_B = 1.2$. Emission from cool gas with temperature $T < 10^7$ K
was neglected because we are interested in time variation of X-ray
luminosity.
\begin{figure}[htbp]
 \caption[]{} \label{tl}
\end{figure}% 

Figure \ref{tlz} plot evolution of $L$ and $L_{\rm n}$ 
against $z$ instead of time.
\begin{figure}[htbp]
 \caption[]{} \label{tlz}
\end{figure}%

In all models the luminosity rapidly rises just before the appearance
of a shock wave, and then rises gradually afterwards. This behavior is
due to the two distinct phases of evolution of gas.
\begin{description}
\item[The first phase (before the emergence of shock wave):] 
Through accretion of ambient gas towards the center gas in the central
region is compressed adiabatically so that temperature and density
there rapidly rise until temperature reaches about virial temperature.
The rapid increase in $L$ is thus due to a rapid rise in
temperature of the central region. 
\item[The second phase (after the appearance of shock wave):] 
As the shock wave propagates outwards, high temperature region
expands, thus increasing $L$.  At the same time, gas inside the
shock wave continues to be compressed adiabatically, thereby its
emissivity being increased gradually.  Luminosity thus increases
faster than $[r_{\rm shock}(t)]^3 \propto t^{2.5}$.
\end{description}

\subsection{Dependence on the Cosmological Density Parameter}

In this subsection we compare the results of Model A ($\Omega_0 =1$)
and Model B ($\Omega_0 = 0.2$) to discuss the $\Omega_0$ dependence of
the cluster evolution under the condition that the perturbation 
amplitudes at $z=10$ and total masses contained in the perturbations
are similar.

From table \ref{dmden}, we see that central DM density,
$\rho_{{\rm DM},0}$,is proportional to $\Omega_0$.  
The DM distribution is less
concentrated in Model B than in Model A, because the ambient gas
density is lower in a lower density universe so that a larger volume
is needed to contain the same amount of mass ($\sim 10^{15} M_{\odot}$) 
initially.  
The absolute values of DM density are thus different among these models,
but the shape of the DM density profiles are similar. This explains why
nondimensional DM density profiles look similar among both models 
(figure \ref{ndden}a).
On the other hand, central density of gas, $n_{{\rm gas},0}$,
depends also on its initial entropy and is, hence,
not strictly proportional to $\Omega_0$ (see table \ref{gasden}).
This is responsible for the
differences in nondimensional gas density profiles among two models
(figure \ref{ndden}b).

In both models X-ray luminosity rises, however, the increase in $L(t)$
is slower in Model B than in Model A. This tendency is more clearly
seen when $L(t)$ is plotted against times, rather than against
redshifts (see Fig. \ref{tl}b, \ref{tlz}b). In the latter figure, the
difference can be recognized at $z \sim 0.5$, but not at $z > 1$.
This is because smaller amount of DM and gas surrounding a central
condensation in a lower $\Omega_0$ universe.

\subsection{Dependence on the Initial Temperature of Gas}\label{ditg}

We calculated a model with lower initial temperature (LT) to see how
initial temperature affects the later evolution.  The temperature and
entropy profiles at $z=0$ of Models A and LT are depicted in figure
\ref{ent}. There is no significant difference among these models,
especially in the structure inside the shock front. The thermal
property of gas outside the shock front has little influence on the
thermal property of gas inside because of enormous entropy production
at the shock surface.

On the other hand, the central densities are different among these
models (figure \ref{denlt}).  According to $n_{{\rm gas},0}$ estimated from the
fitting data of table \ref{gasden} and figure \ref{denlt}, central
density of Model LT is about three times greater than that of Model A. 
In the density profiles at $r >0.3$ Mpc, however, any difference can 
hardly be seen.

\begin{figure}[htbp]
 \caption[]{} \label{denlt}
\end{figure}%

In the central region, gas is adiabatically compressed until the
temperature reaches about the virial temperature. So lower initial gas
temperature results in higher gas density, if the virial temperatures
are the same in both models. However, the difference in central gas 
densities cannot be perfectly explained by this picture. If there
is no entropy production (i.e., gas is perfectly adiabatically compressed),
central density of Model LT should be about thirty times greater than 
that of Model A.  This means, shock heating took away the information 
about the initial gas temperature also in the central region.

Because of higher central gas density Model LT cluster is more luminous in
X-ray than Model A cluster (figure \ref{tl}a, \ref{tlz}b). Substantial 
difference, however, cannot be seen in evolutionary behavior of X-ray
luminosity of both model.  (figure \ref{tl}b, \ref{tlz}b).

\section{SUMMARY AND DISCUSSION}\label{summary}

We carry out the numerical simulations of spherical clusters of
galaxies with shell model for DM and second order up-wind TVD scheme
for ICM to examine structural evolution of ICM.  Shock front
moves outwards as gas accretes towards cluster center, yielding a
relatively flat temperature profile inside the shock front. Density
and pressure profiles evolves in a self-similar fashion.

X-ray luminosity increases with time in two steps. 
At the initial collapse of DM
gas in the central part is at first adiabatically compressed through 
accretion of ambient gas towards the center.  Eventually,
a shock wave appears near the
center and X-ray luminosity rapidly rises until temperature increases
and reaches about virial temperature via shock heating.  
In the late stage, in contrast, the luminosity rises only gradually, since the
inner region already emits strong X-rays.  The gradual brightening is
due partly to the expansion of the high temperature region and partly
to increasing X-ray emissivity of gas as the results of continuous
adiabatic compression the gas inside the shock.

If we compare two clusters with the same density fluctuation amplitudes
at $z=10$ and with the same total masses but in different mean-density
unvierses,
the DM distribution is less concentrated in clusters in lower density
universe.  Hence, X-ray luminosity of clusters rises more slowly
than in higher density universe.

Initial gas temperature has some influence on the central gas density.
Higher initial temperature results in lower central density.
This corresponds to the case that reheating of the ICM by, e.g.,
proto-galaxies is substantial.  Density distribution in the outer
region and temperature profile inside the shock front are, however, 
hardly influenced by changes in the initial temperature.  Thus, the
inclusion of reheating process modifies the scaling law between X-ray
luminosity and temperature in the favorite way to reproduce the
observed relation (NFW).  Note that the epoch of
the shock emergence depends on the initial specific entropy at the core;
in the presence of reheating process (so that the initial gas temperature
is $\sim 10^7$K), specific entropy at the core is already high enough,
and so the appearance of shock may be delayed.

In table \ref{bertschinger} we compare the time dependent properties
of Model A cluster and those of the self-similar solution by
Bertschinger (1985).  Both behavior looks very similar.  Note that the
self-similarity can be seen in all the calculated models.  Although
Einstein-de Sitter universe is assumed in Bertschinger (1985), we find
that the self-similarity can be found for cluster evolution in low
density universe ($\Omega_0 = 0.2, \lambda_0 = 0$).

Our results regarding the present profiles of density, temperature,
and so on roughly coincide with those of the previous SPH simulations
(Evrard 1990; NFW). 
However, the structure of the
shock front, which was not well resolved in the previous SPH simulations, 
is now clearly captured in the present mesh-code simulations; our results
show that hot gas of X-ray CG is separated with a definite boundary
and that shock heating plays an important role for the heating process
of ICM. We confirmed the persistence of radial infall of gas after 
the passage of the shock front. This results in adiabatic compression
of the inner parts, inducing gradual temperature increase.
We also found time variations in the strengths of the shock due to
a sound wave propagation over the entire cluster, which modulates
the radial distributions of temperature and density with
relative amplitudes of about 10 \%.
Note that the property of the
sound wave may depend on our assumption of spherical symmetry.
It is open to question how such sound wave behaves in a
realistic three-dimensional situation; i.e, when mass accretion takes
place in a nonaxisymmetric way (e.g., by merger of multiple density
condensations). 
If the temperature modulation will be observed with
future X-ray mission (such as ASTRO-E), this can be used as a
good probe to investigate the structure, especially, the mass
distribution of CGs, since sound wave properties sensitively depend on
the shape and the depth of the gravitational potential well,
and thus on the DM mass distribution.

\acknowledgements

We would like to thank F.E.Nakamura and M.Hattori for helpful 
discussions and comments. This
work is in part supported by Research Fellowships of the Japan Society
for the Promotion of Science for Young Scientists (M.T.), and by the
Grants-in Aid of the Ministry of Education, Science, Sports, and
Culture of Japan, 06233101, 08640329 (S.M.).

\clearpage
 \begin{table}
  \begin{center}
   \begin{tabular}{ccccc} 
    \hline \hline
    Model  & $\delta_0$ & $r_0$ (Mpc) & $T_{\rm gas,i}$ (K) & $\Omega_0$ \\
    \hline
    A      & $0.4$      & $9.0$        & $10^7$        & $1.0$      \\
    B      & $0.5$      & $15.4$       & $10^7$        & $0.2$      \\
    LT     & $0.4$      & $9.0$        & $10^6$        & $1.0$      \\
    \hline
   \end{tabular}
  \end{center}
  \caption[Parameters of each model.]{Parameters of each model.}
  \label{parameters}
 \end{table}%

\clearpage
 \begin{table}
  \begin{center}
   \begin{tabular}{ccccc} 
    \hline \hline
    Model  & $\rho_{{\rm DM},0}$ ($10^{-26}$ g cm$^{-3}$) & $r_{\rm DM}$ (Mpc) & $\beta_{\rm DM}$ & $r_{100}$  \\ 
    \hline
    A      & $78.5$   & $0.067$  & $0.87$ & $1.35$  \\
    B      & $14.9$    & $0.13$   & $0.93$ & $1.04$  \\
    LT     & $83.3$   & $0.070$  & $0.87$ & $1.34$  \\
    \hline
   \end{tabular}
  \end{center}
  \caption[Density profiles of DM at $z=0$.]{Density profiles of DM at $z=0$.}
  \label{dmden}
 \end{table}%

\clearpage
 \begin{table}
  \begin{center}
   \begin{tabular}{cccc} 
    \hline \hline
    Model  & $n_{{\rm gas},0}$ ($10^{-3}$cm$^{-3}$) & $r_{\rm gas}$ ({\rm Mpc}) &  $\beta_{\rm gas}$ \\ 
    \hline
    A      & $9.9$      & $0.082$         &  $0.80$        \\
    B      & $3.9$      & $0.13$          &  $0.93$        \\
    LT     & $30$      & $0.065$         &  $0.87$        \\
    \hline
   \end{tabular}
  \end{center}
  \caption[Density profiles of gas at $z=0$.]{Density profiles of gas at $z=0$.}
  \label{gasden}
 \end{table}%

\clearpage
 \begin{table}
  \begin{center}
   \begin{tabular}{ccc} 
    \hline \hline
    Model  & $r_a$ ({\rm Mpc}) & $v_{a}$ (km s$^{-1}$)  \\ 
    \hline
    A  & $1.2$       & $210$             \\
    B  & $1.4$       & $230$             \\
    LT & $1.2$       & $230$             \\
    \hline
   \end{tabular}
  \end{center}
  \caption[The arrival radius of shock surface at $z=0$, $r_a$, and mean propagating velocity, $v_a$.]{The arrival radius of shock surface at $z=0$, $r_a$, and mean propagating velocity, $v_a$.}
  \label{shockpara}
 \end{table}%

\clearpage
 \begin{table}
  \begin{center}
   \begin{tabular}{ccccc}
    \hline \hline
              & $\rho_{\rm DM}(r)$ & $\rho_{\rm gas}(r)$ & $P(r)$  &  $r_{\rm shock}(t)$ \\
    \hline
    A         & $r^{-2.6}$     & $r^{-2.4}$      & $r^{-3.0}$ & $t^{0.83}$  \\ 
Bertschinger(1985)  & $r^{-2.25}$ & $r^{-2.4}$           & $r^{-2.9}$  & $t^{0.89}$   \\ 
    \hline
   \end{tabular}
  \end{center} 
  \caption[The comparison between the behavior of A and the self-similar solution of Bertschinger (1985) ]{The comparison between the behavior of A1 and the self-similar solution of Bertschinger (1985) }
 \label{bertschinger}
 \end{table}%

\clearpage
\begin{center}
{\bf Figure Captions}
\end{center}

\begin{description}
\item[figure \ref{indmden}]
     (a) Radial profiles of initial density perturbations and 
     (b) $(\delta M / M)/\sigma_{8,{\rm CDM}}$,
     where $(\delta M / M)$ is the mass fluctuation averaged
     over a scale $r$ 
     and $\sigma_{8,{\rm CDM}}$ is
     the rms of mass fluctuation of scale $r$ obtained from CDM power spectrum
     with $\sigma_8$ being a normalization.  The solid line correspond to 
     Model A and the dotted line corresponds model B.
\item[figure \ref{dms8o1.0ex}] The evolution of (a) density profile and 
      (b) radial velocity profile of DM of Model A cluster.
      The different lines correspond to different redshifts (and different 
      times): $z=2.5$ ($t=1$ Gyr) by the long dash line; $z=1.2$ ($t=2$ Gyr)
      by the short dash line; $z=0.4$ ($t=4$ Gyr) by the dotted line; 
      and $z=0$ ($t=6.5$ Gyr) by the solid line, respectively.
\item[figure \ref{gascla1}] Evolution of Model A cluster.  Radial profiles 
      of representative physical quantities of gas are depicted:
     (a) radial velocity, (b) density, (c) temperature, (d) pressure, and
     (e) entropy. The different types of lines correspond to same redshifts as
      in figure \ref{dms8o1.0ex}
\item[figure \ref{ndden}] Nondimensional density profiles at $z=0$ for 
     Models A (by the solid line) and for B (by the dotted line), respectively.
     Here, density is scaled with $\rho_{{\rm c}0}$, critical density of 
     the universe at $z=0$, and radius is scaled with $r_{100}$.
\item[figure \ref{ent} ] (a) Temperature profiles and (b) entropy profiles 
     of each model at $z=0$. Entropy is normalized to zero at the outer 
     boundary.  The different types of lines correspond to different
     models: Model A by the solid line; Model B by the dotted line; 
     and Model LT by the short dash line, respectively.
\item[figure \ref{tsuriai}] The ratio of pressure gradient force and
      gravitational force inside the shock surface of Model A cluster
      from $t=5.5$ Gyr to $6.5$ Gyr.
      %where $|dP/dr|$ is the absolute value of pressure gradient
      %and $F_{\rm g} \equiv G M_{r} \rho_{\rm gas} / r^2$. 
      If the system is perfectly in hydrostatic equilibrium, 
      the ratio, $|dP/dr|/F_{\rm g}$, should be unity.
\item[figure \ref{wave}]  Radial velocity of gas inside shock front 
      from $t=5.5$ Gyr to $6.5$ Gyr of model A. Sound wave propagates 
      outwardly. Radial infall systematically remains even after the 
      passage of the shock.
\item[figure \ref{shock} ] Time evolution of the radius of a shock surface,
      $r_{\rm shock}$.  The abscissa is the look-back time, time measured
      from the present time; $t_{{\rm lb}}=0$ at $z=0$.
      Shock surface moves with a nearly constant velocity 
      ($\sim 200$ km s$^{-1}$).  The different types of lines correspond 
      to same models as in figure \ref{ent}.
\item[figure \ref{shockamp} ]  Evolution of shock strength and 
      the shock radius of Model A. (a) Time evolution of $r_{{\rm shock}}$
      by the dotted line and that of $\Delta v$,  
      jump in the radial velocity of gas over the shock surface.
      (b) That of $T_7 \Delta S$, where $T_7$ is 
      the pre-shock gas temperature in the
      unit of $10^7 {\rm K}$ and $\Delta S$ is the jump in the entropy over 
      the shock surface. This quantity is proportional to heat produced through
      shock heating.
\item[figure \ref{tl} ] (a) Time evolution of X-ray luminosity, 
      $L(t_{\rm lb})$, and (b) that of normalized luminosity, 
      $L_{\rm n}(t_{\rm lb}) \equiv L(t_{\rm lb})/L(t_{\rm lb}=0)$.
      %which is the luminosity normalized with the present value. 
      In all models $L$ rapidly rises
      just before the generation of a shock wave, and then rises gradually
      afterwards.  The different types of lines
      correspond to same models as in figure \ref{ent}.
\item[figure \ref{tlz} ] Evolution of $L$ and $L_{\rm n}$ plotted
      against $z$.
\item[figure \ref{denlt} ] Gas density profiles for Model A (solid line)
      and for Model LT (dotted line).  Central density is about 
      three times greater in Model LT cluster than in Model A cluster.  In 
      the region of $r >0.3$Mpc, however, this difference can hardly be seen. 

\end{description}


\begin{references}
\reference{} Anninos, P., \& Norman, M. L. 1996, ApJ, 459. 12
\reference{} Bertschinger, E. 1985, ApJS, 58, 39
\reference{} Binney, J., \& Cowie, L. 1981, ApJ, 247, 464
\reference{} Bower, R. G., B\"{o}hringer, H., Briel, U. G., Ellis, R. S., Castander, F. J., \& Couch, W. J. 1994, MNRAS, 268, 345
\reference{} Bryan, G. L., Cen, R., Norman, M. L., Ostriker, J. P., \& Stone, J. M. 1994, ApJ, 428, 405
\reference{} Castander, F. J., Ellis, R. S., Frenk, C. S., Dressler, A., \& Gunn, J. E. 1994, ApJ, 424, L79
\reference{} Cen, R. \& Ostriker, J. P. 1994, ApJ, 429, 4
\reference{} Edge, A. C., Stewart, G. C., Fabian, A. C., \& Arnaud, K. A. 1990, MNRAS, 245, 559
\reference{} Evrard, A. E. 1990, ApJ, 363, 349
\reference{} Fabian, A. C. 1994, ARA\&A, 32, 277
\reference{} Fabricant, D. G., Kent, S. M. \& Kurtz, M. J. 1989, ApJ, 336, 77
\reference{} Gioia, I. M., Henry, J. P., Maccacaro, T., Morris, S. L., Stock,J.T., \& Wolter,A. 1990, ApJ, 356, L35
\reference{} Funato, Y., Makino, J. \& Ebisuzaki, T. 1992a, PASJ, 44, 291
\reference{} Funato, Y., Makino, J. \& Ebisuzaki, T. 1992b, PASJ, 44, 613
\reference{} Gunn, J. E., Gott, J. R. 1973, ApJ, 176, 1
\reference{} Henry, J. P., Gioia, I. M., Maccacaro, T., Morris, S. L., Stocke, J. T., \& Wolter, A. 1992, ApJ, 386, 408
\reference{} H\'enon, M. 1964, Ann. Astrophys., 27, 83
\reference{} Kang, H., Cen, R., Ostriker, J. P., \& Ryu, D. 1994, ApJ, 428, 1
\reference{} Kneib, J. D. P., Mellier, Y., Pello, R., Miralda-Escude, J., Le Borgne, J. F., Boehringer, H., \& Picat, J. P.
\reference{} Lynden-Bell, D. 1967, MNRAS, 136, 101
\reference{} Metzler, C. A., \& Evrard, A. E. 1994, ApJ, 437, 564
\reference{} Monaghan, J. J. 1992,  ARA\&A, 30, 543
\reference{} Nakamura, S. 1996 Doctor Thesis of Tokyo Institute of Technology
\reference{} Navarro, J. F., Frenk, C. S., \& White, S. D. M. 1995, MNRAS, 275, 720
\reference{} Hirsch, C. 1990, Numerical Computation of Internal and External Flows Vol.2:Computational Methods for Inviscid and Viscous Flows (John Wiley \& Sons) 
\reference{} Ohashi et al 1996, X-ray Imaging and Spectroscopy of Cosmic Hot
Plasmas (ASCA 3rd Anniversary)
\reference{} Pistinner, S., \& Shaviv, G. 1996, ApJ, 459, 147
\reference{} Peebles, P. J. E. 1982, ApJ, 257, 438
\reference{} Perrenod, S. C. 1978, ApJ, 226, 566
\reference{} Rosner, R., \& Tucker, W. H. 1989, ApJ, 338, 761
\reference{} Rybicki, G. B., \& Lightman. A. P. 1979, Radiative Process in Astrophysics (John Wiley \& Sons ) 
\reference{} Sarazin, C. L. 1988, X-ray Emission from Clusters of Galaxies (Cambridge University Press)
\reference{} Spitzer, L., Jr. 1962, Physics  of Fully Ionized Gases (New York:Wiley)
\reference{} Suto, Y. 1993, Prog. Theor. Phys., 90, 1173
\reference{} Takizawa, M., \& Inagaki, S. 1996, submitted to PASJ 
\reference{} Thomas, P. A., \& Couchman, H. M. P. 1992, MNRAS, 257, 11
\end{references}
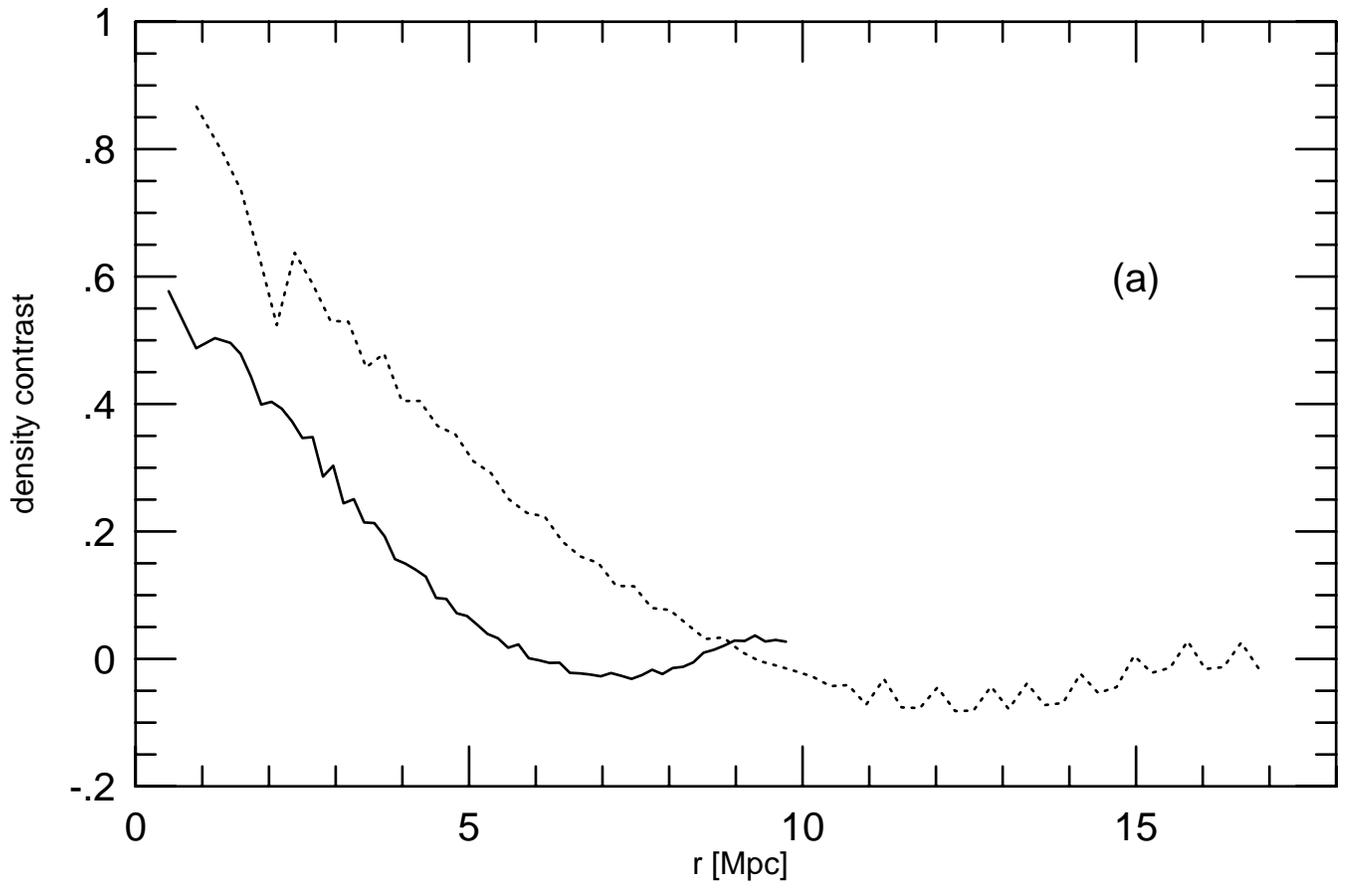
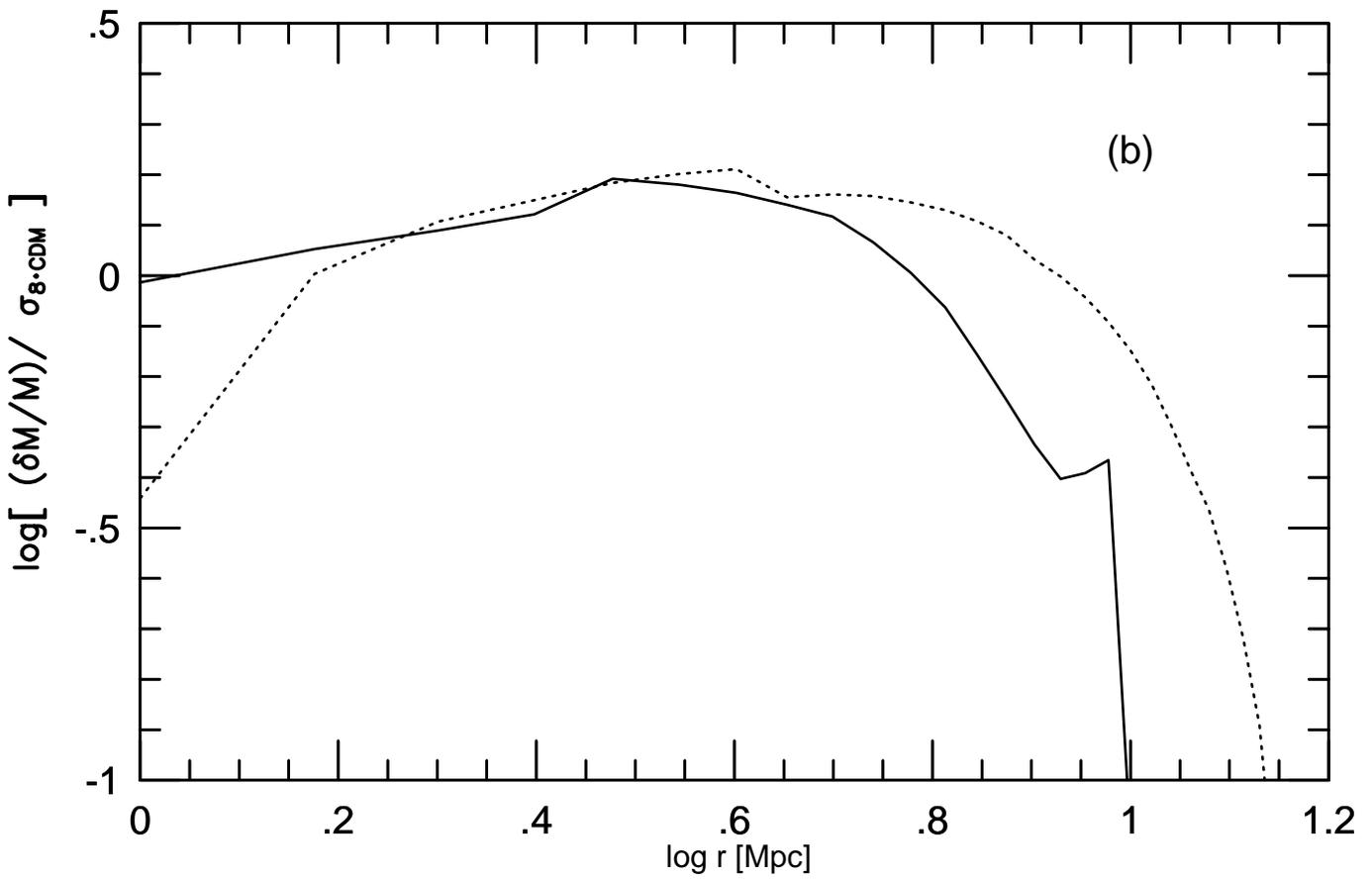
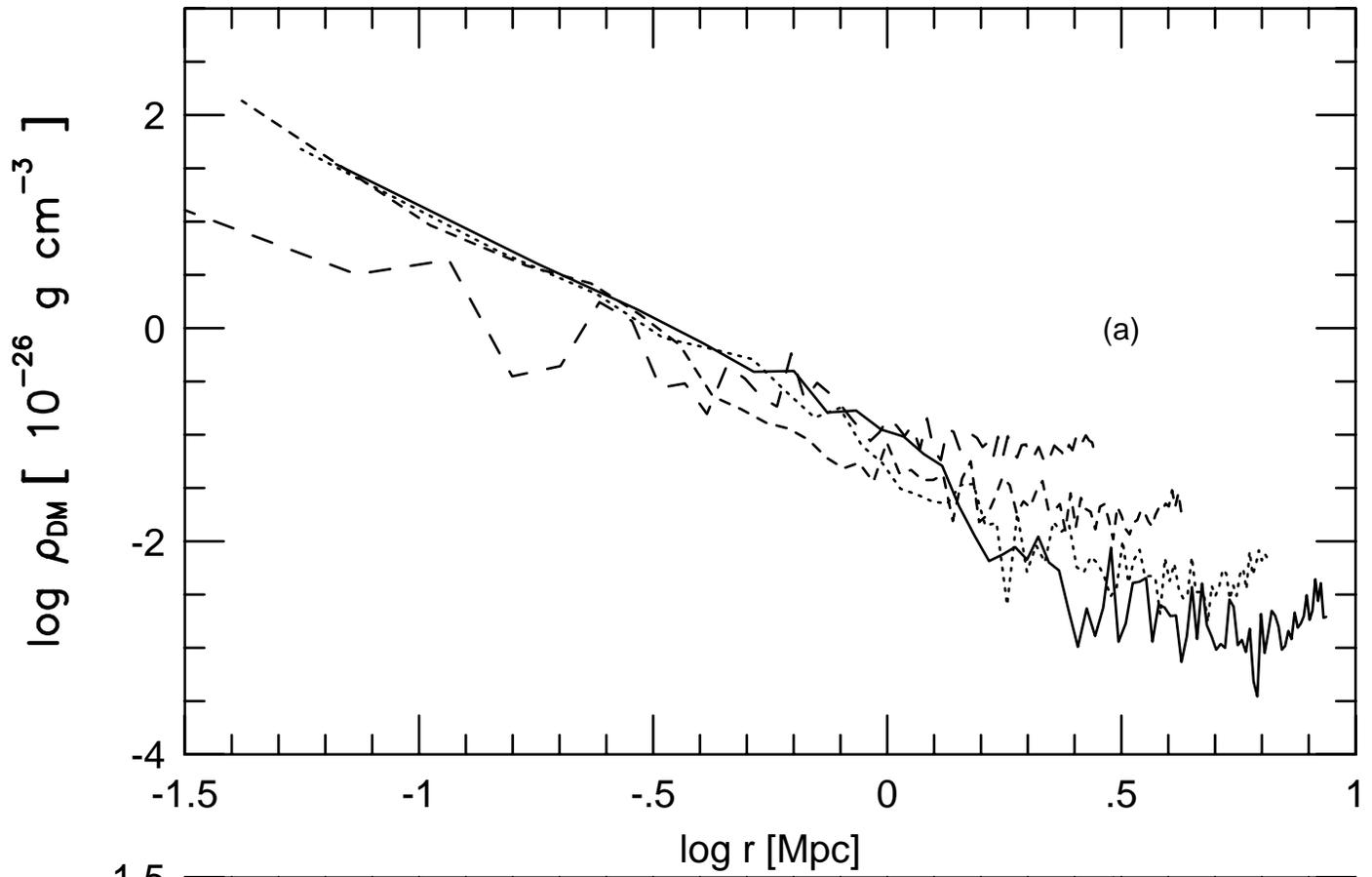
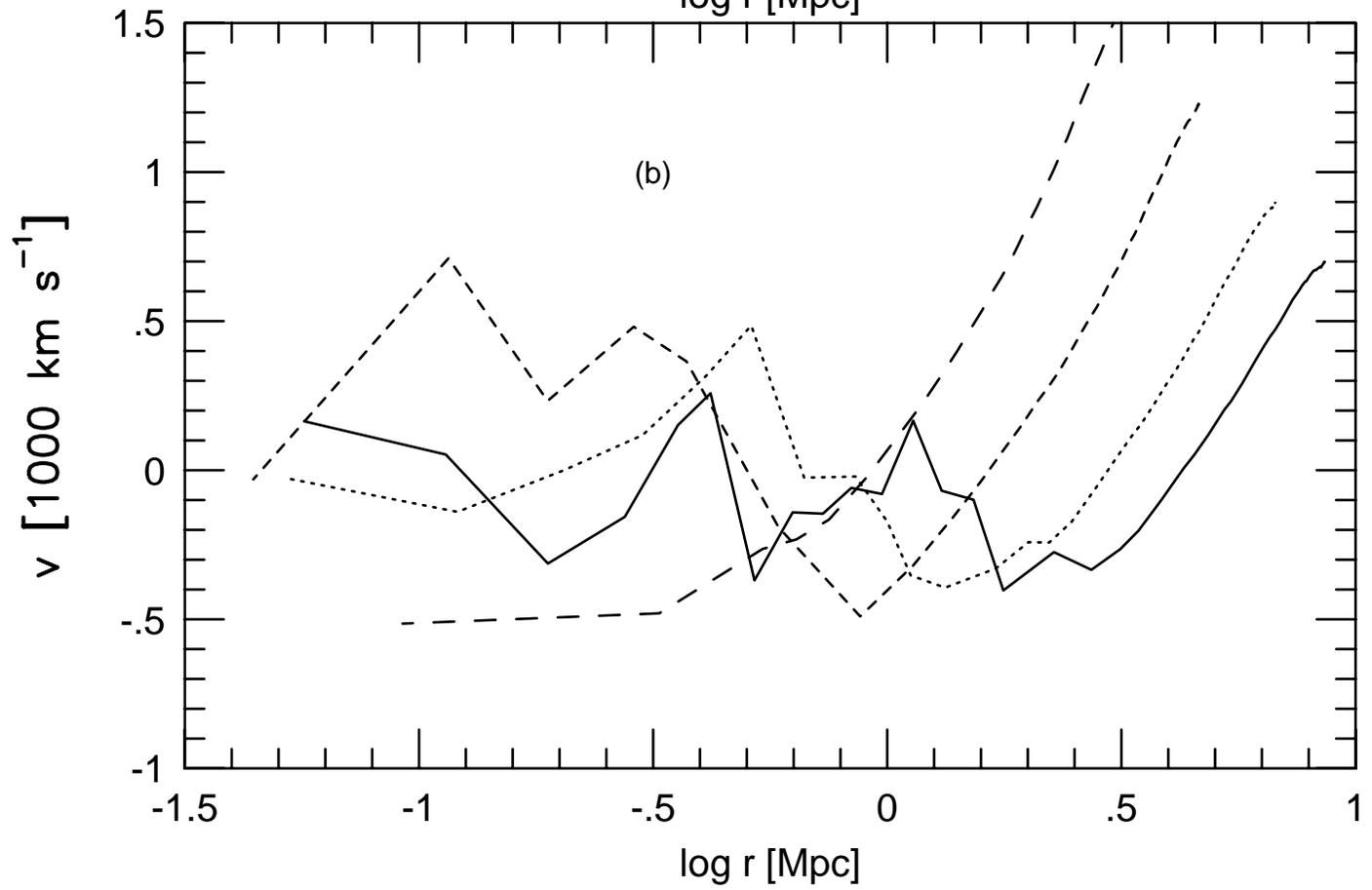
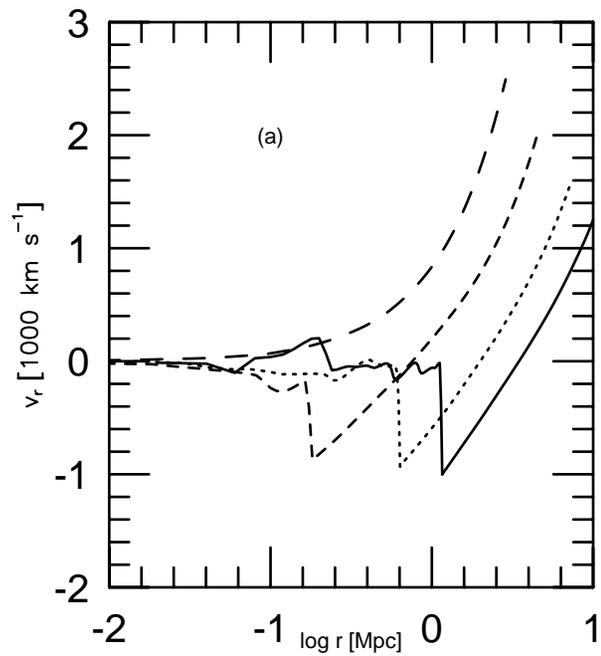
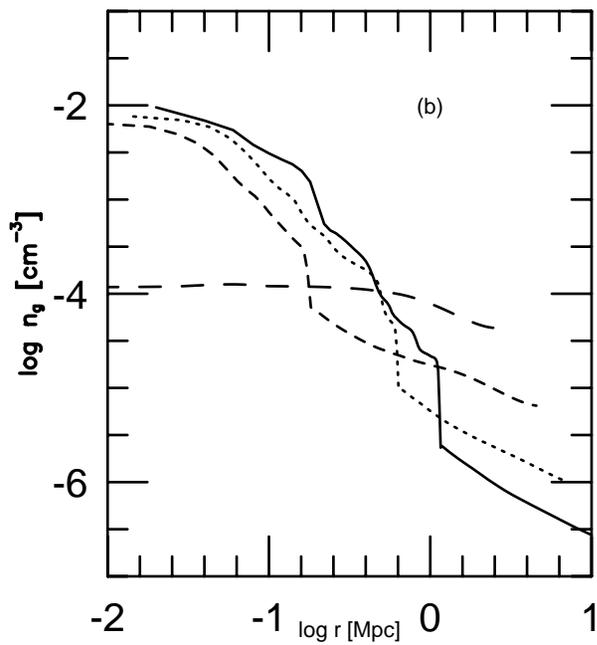
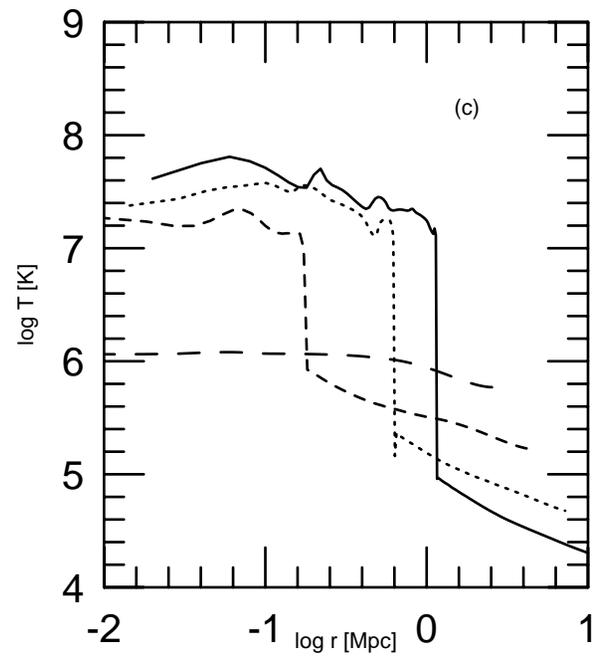
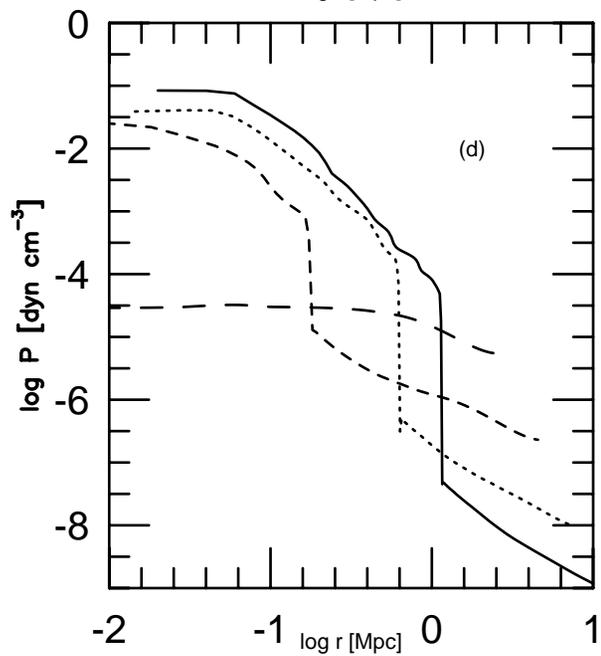
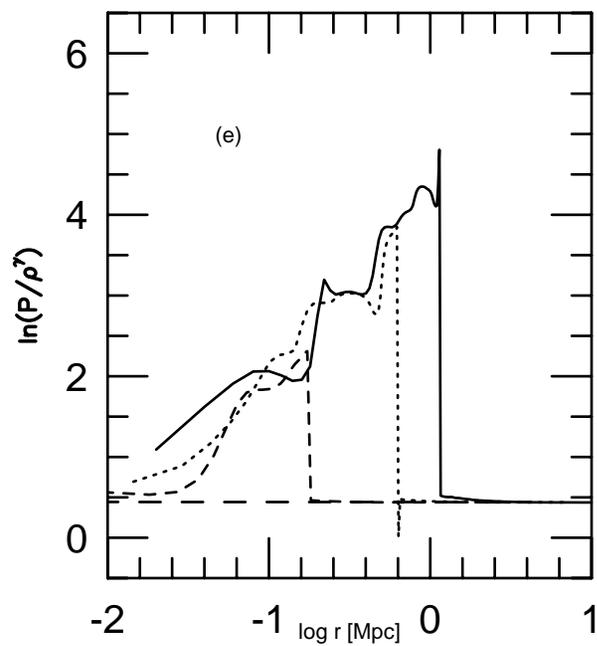
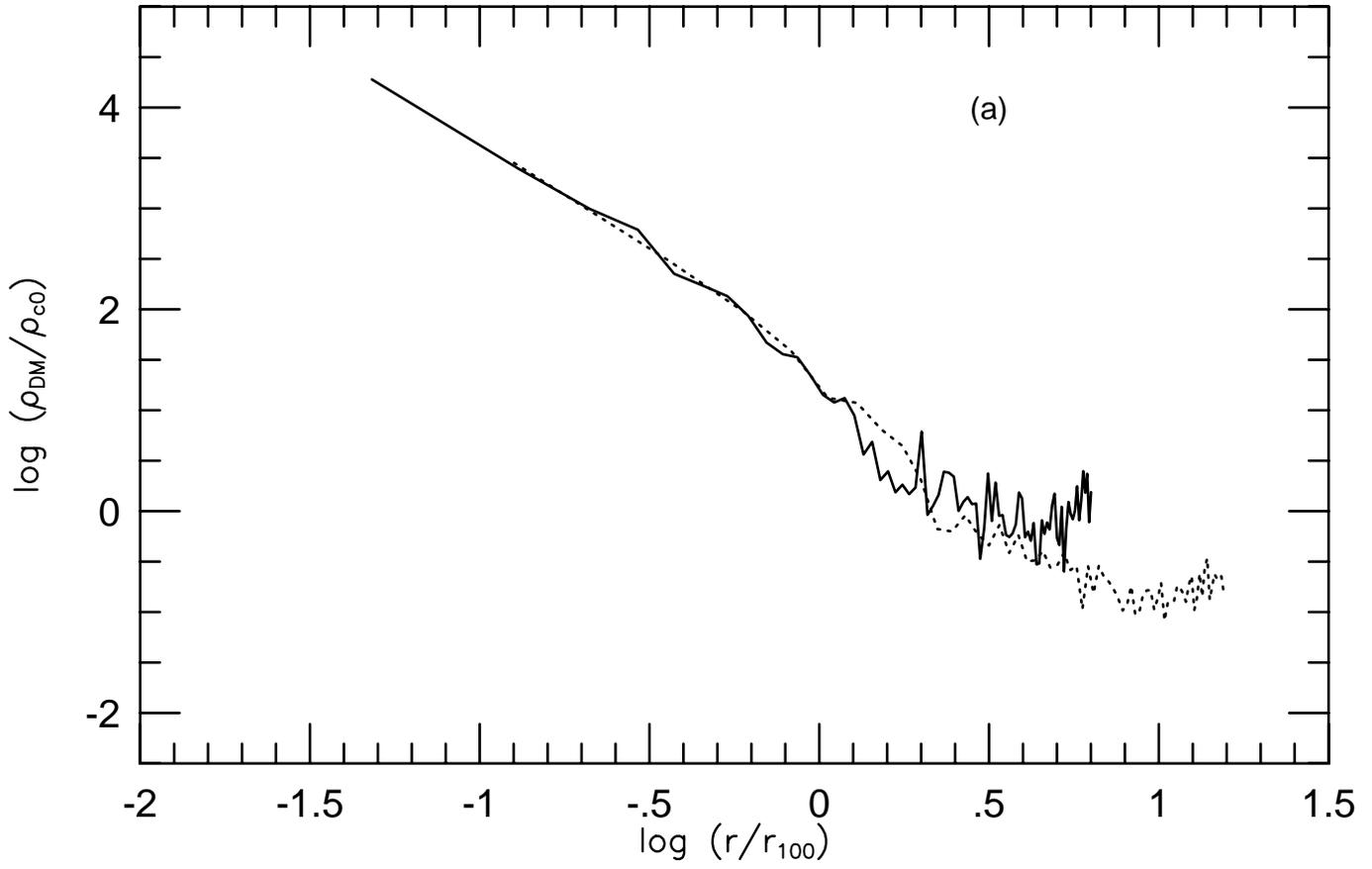
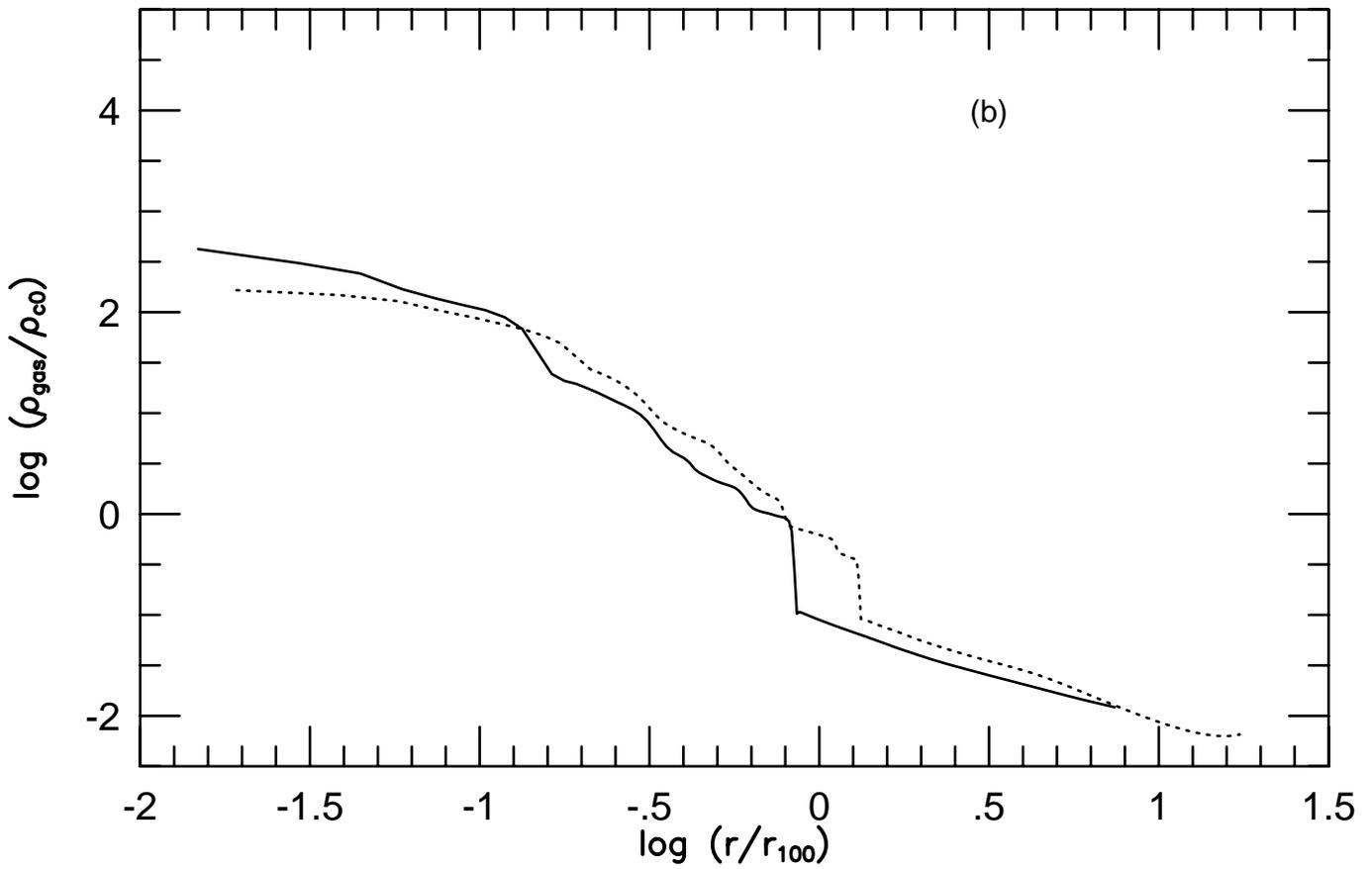
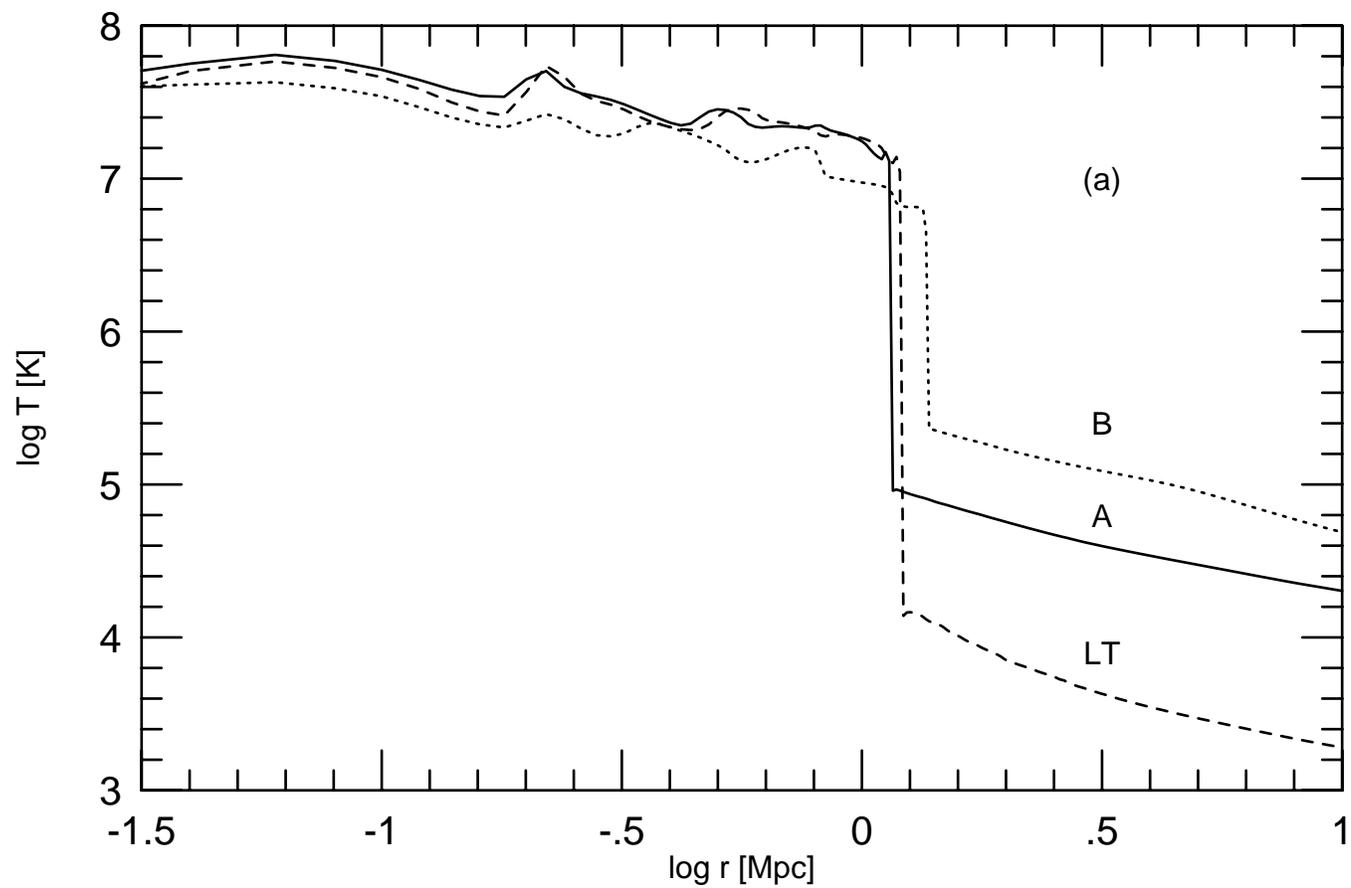
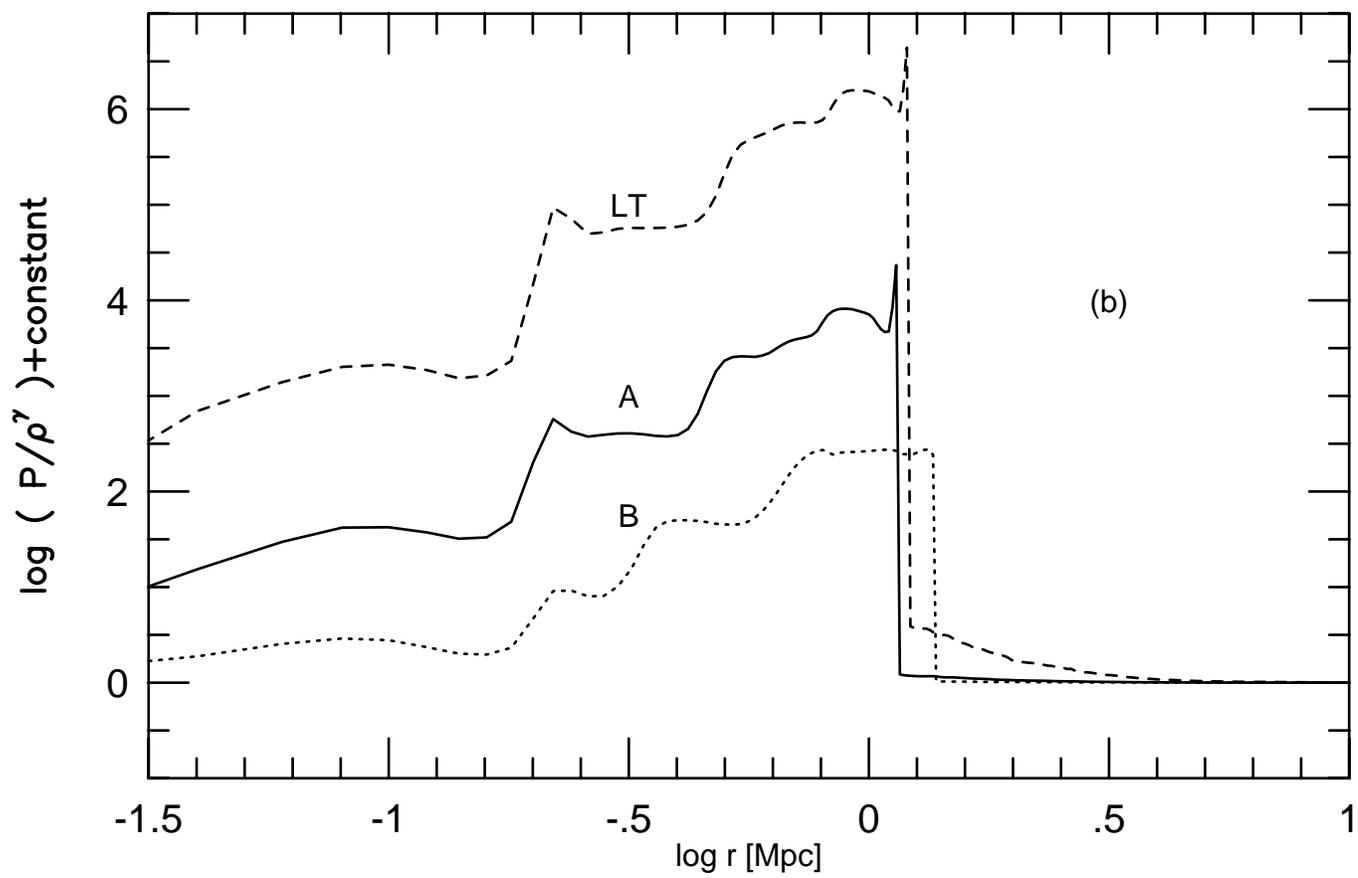
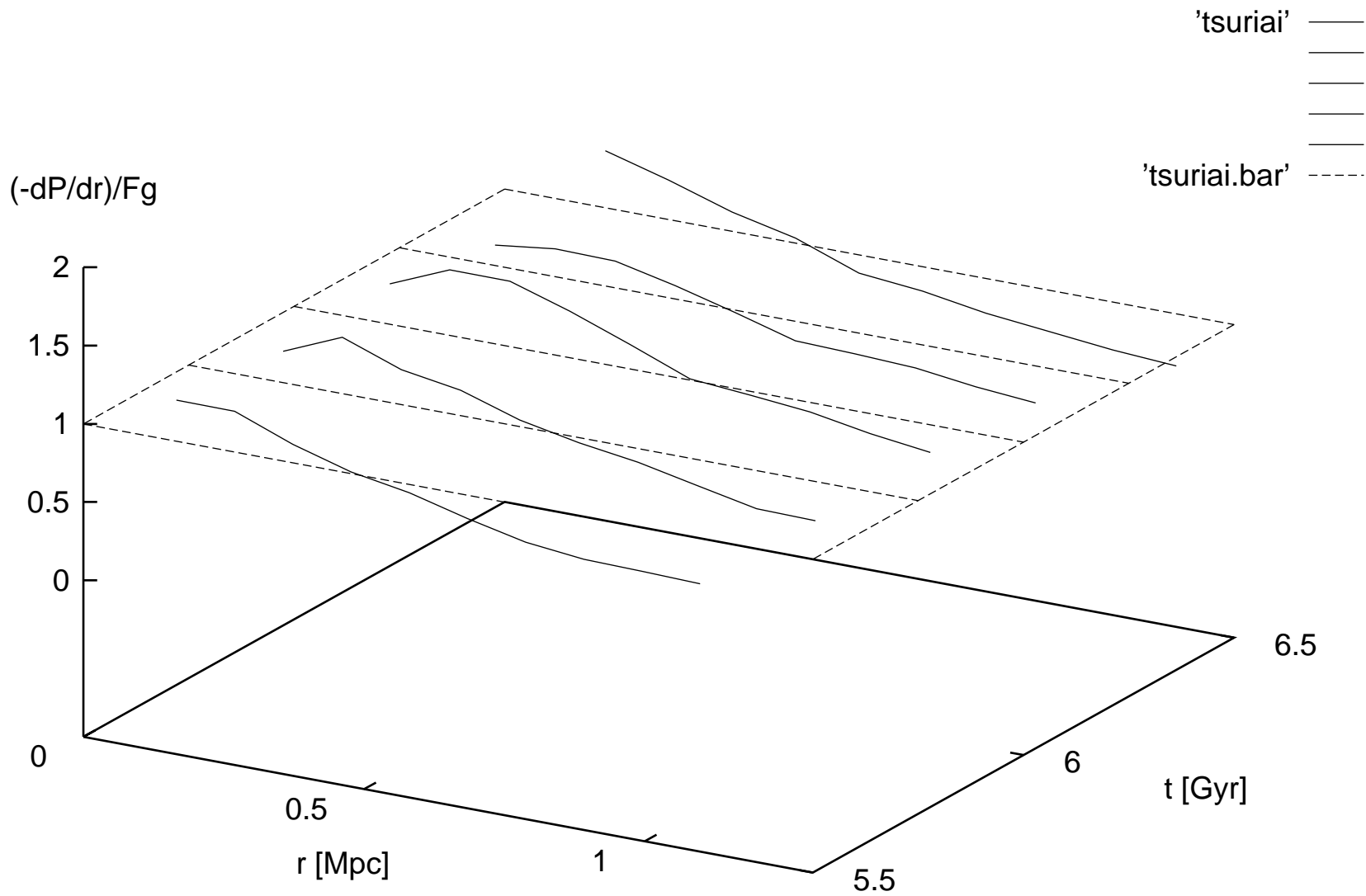
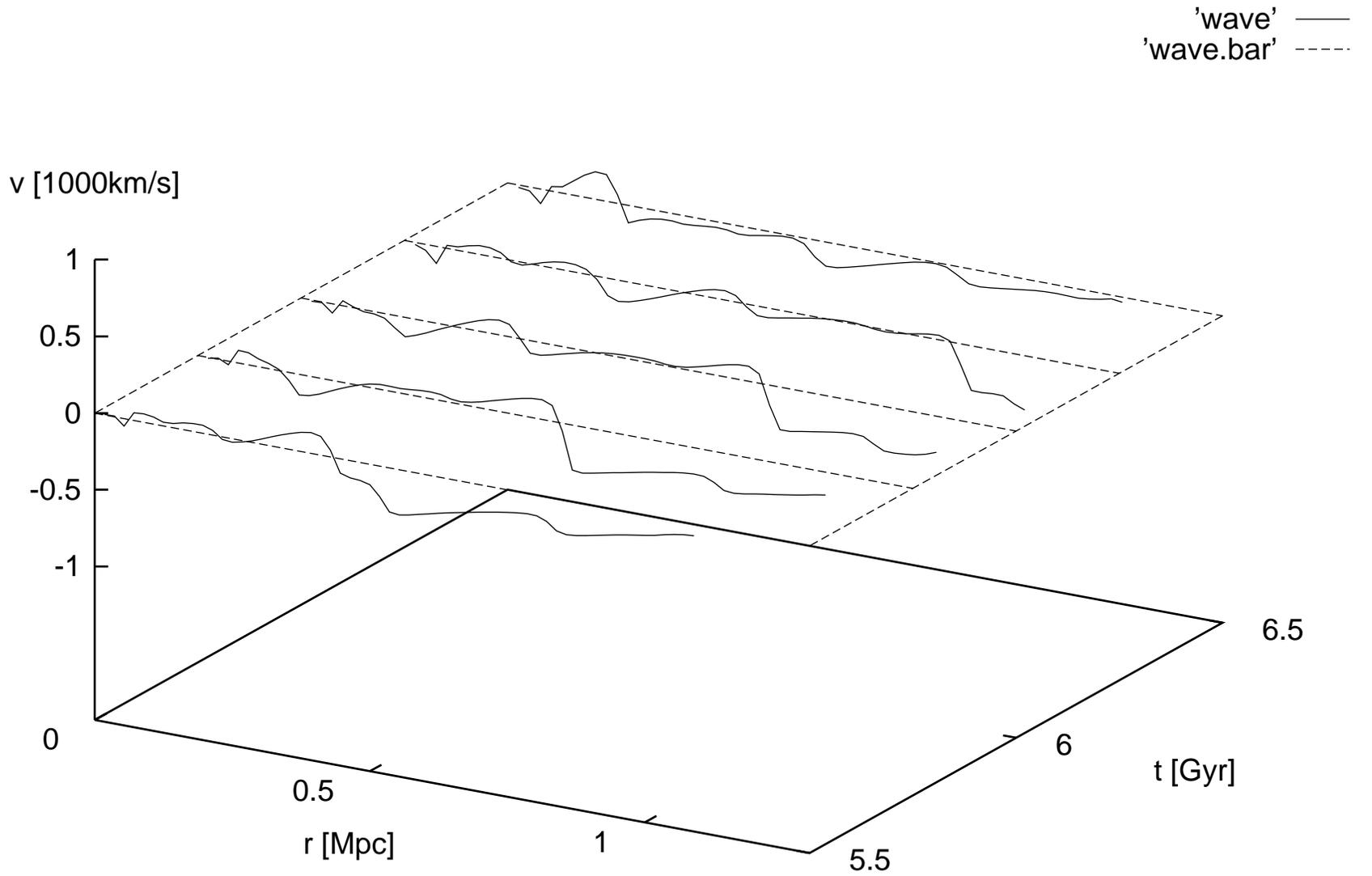
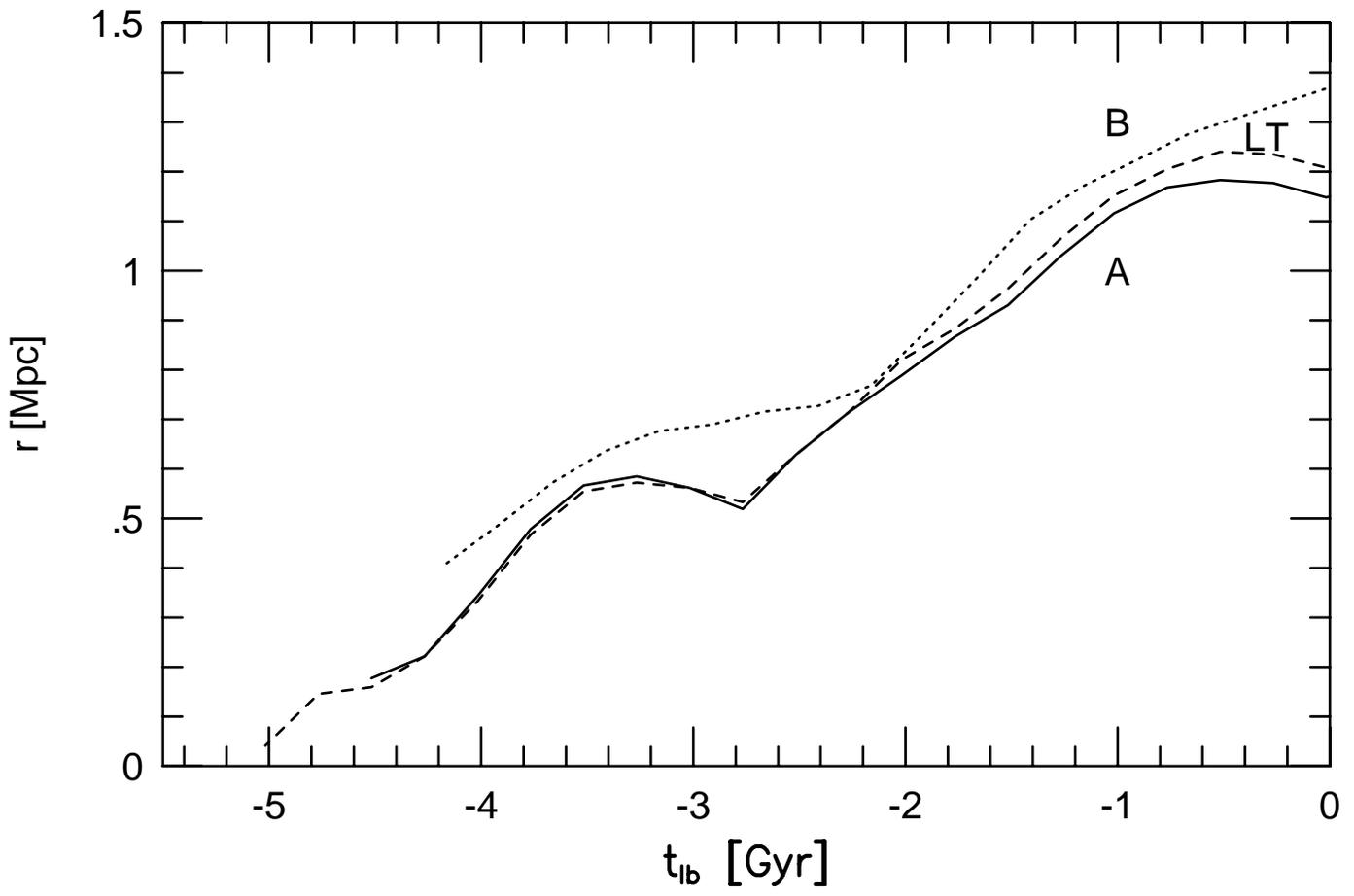
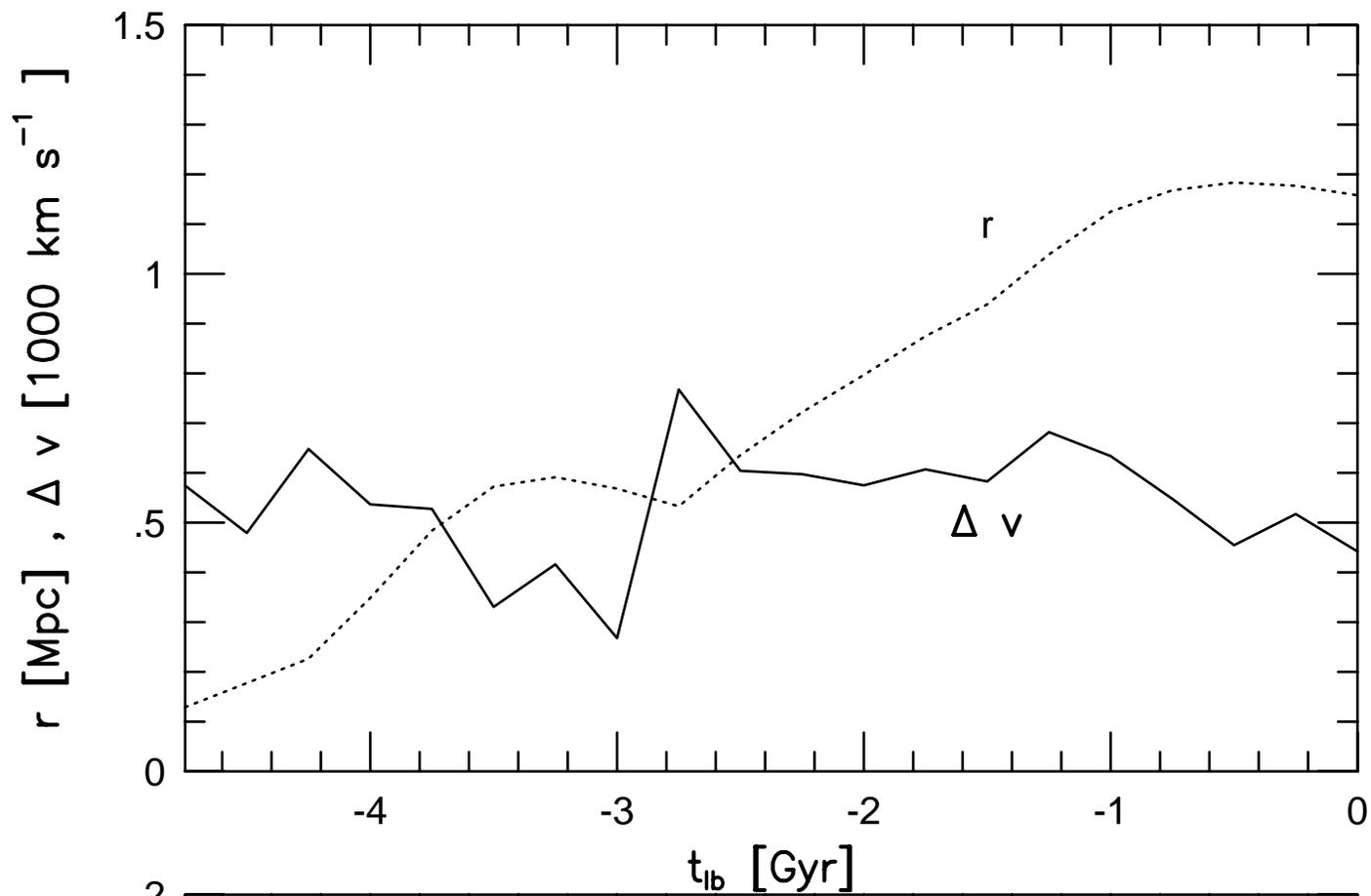
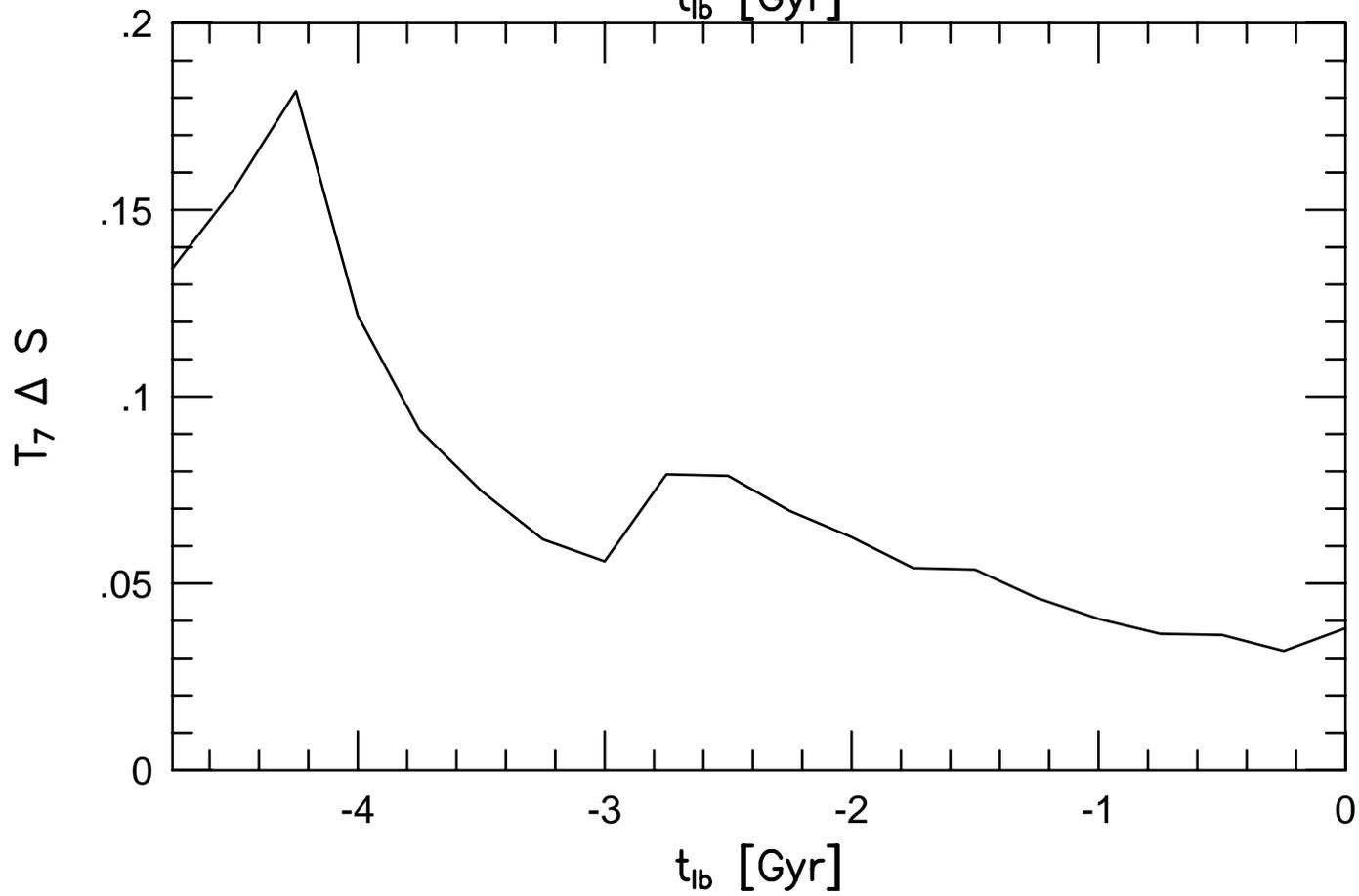
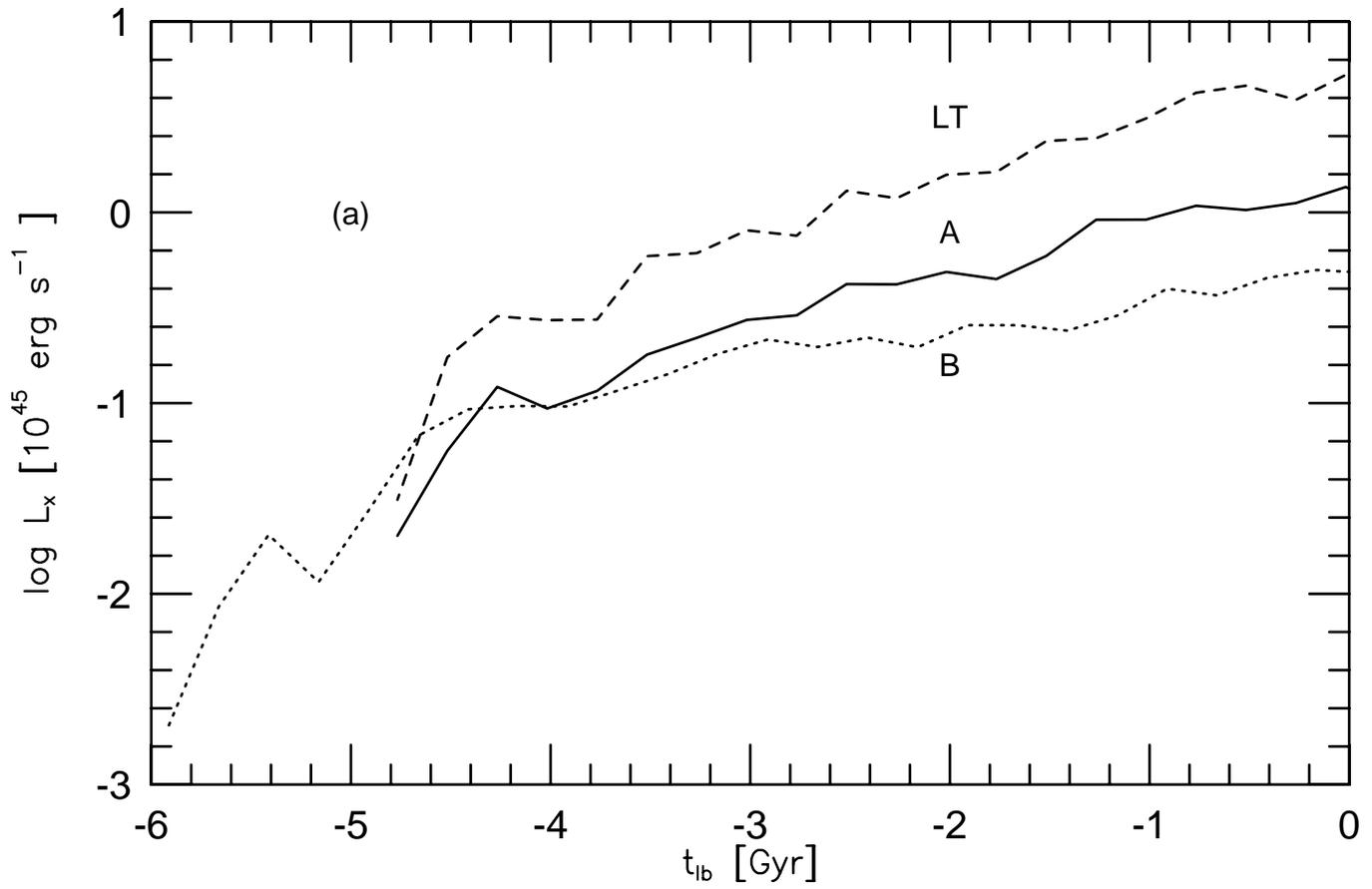
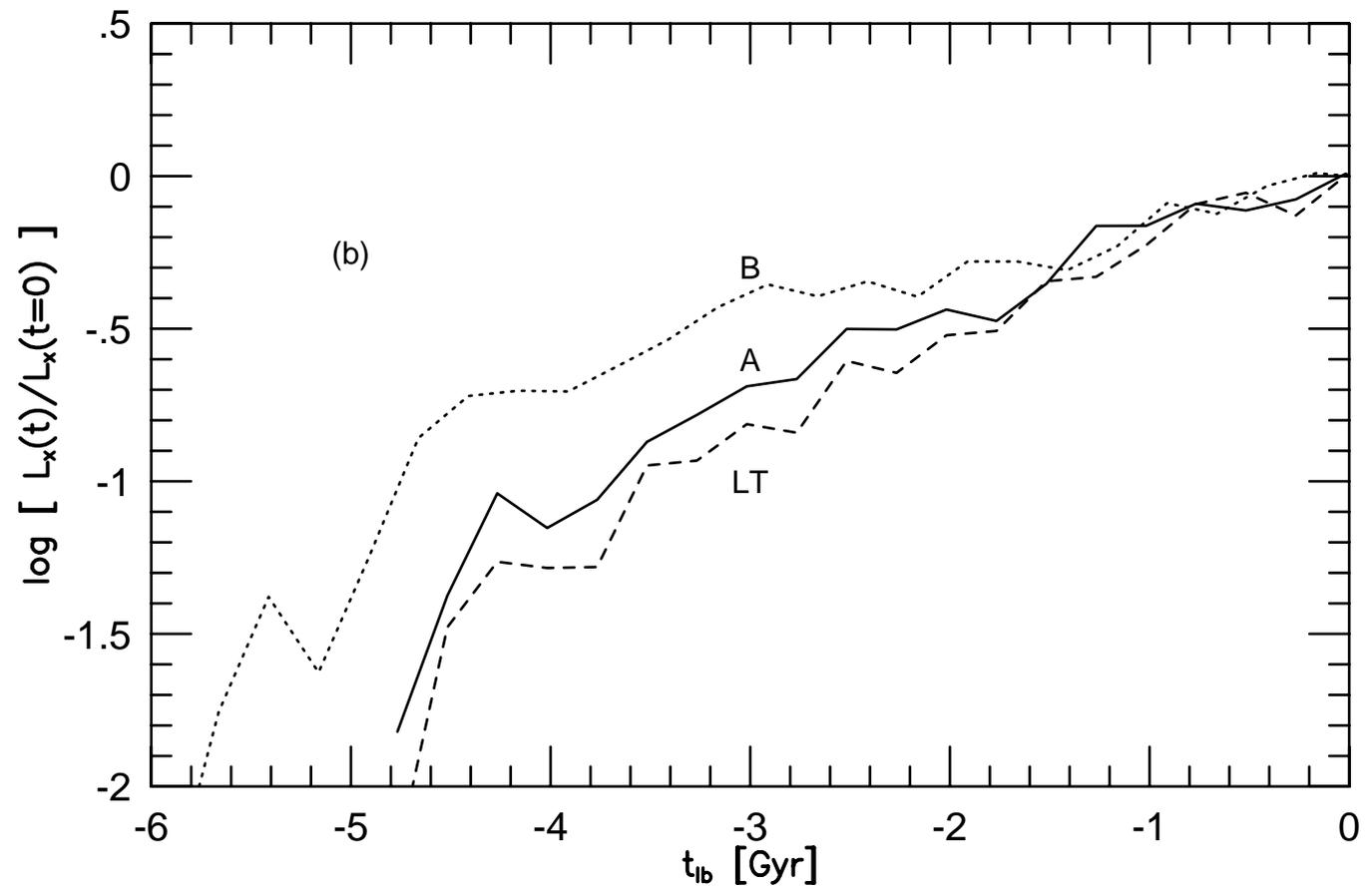
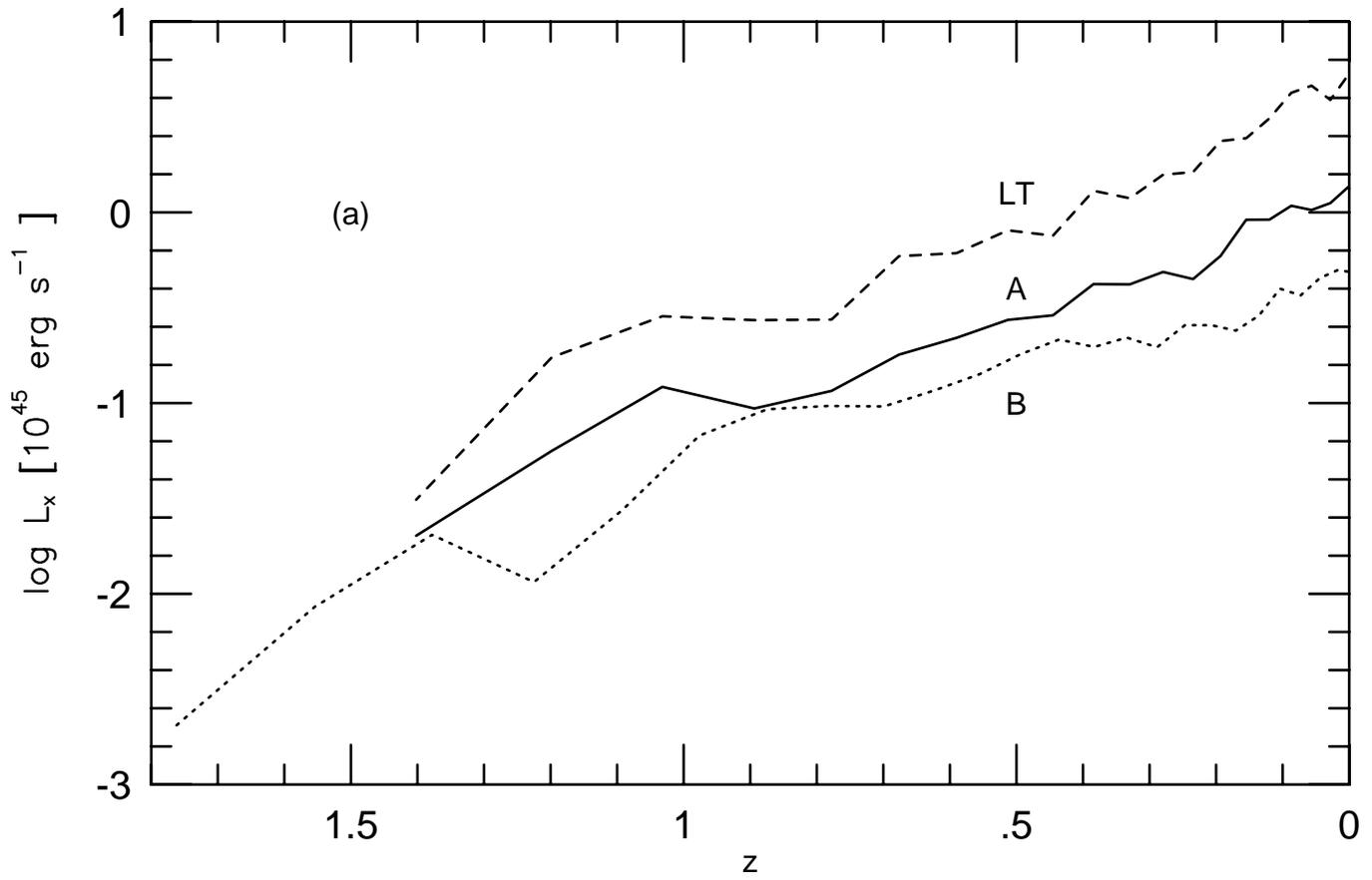
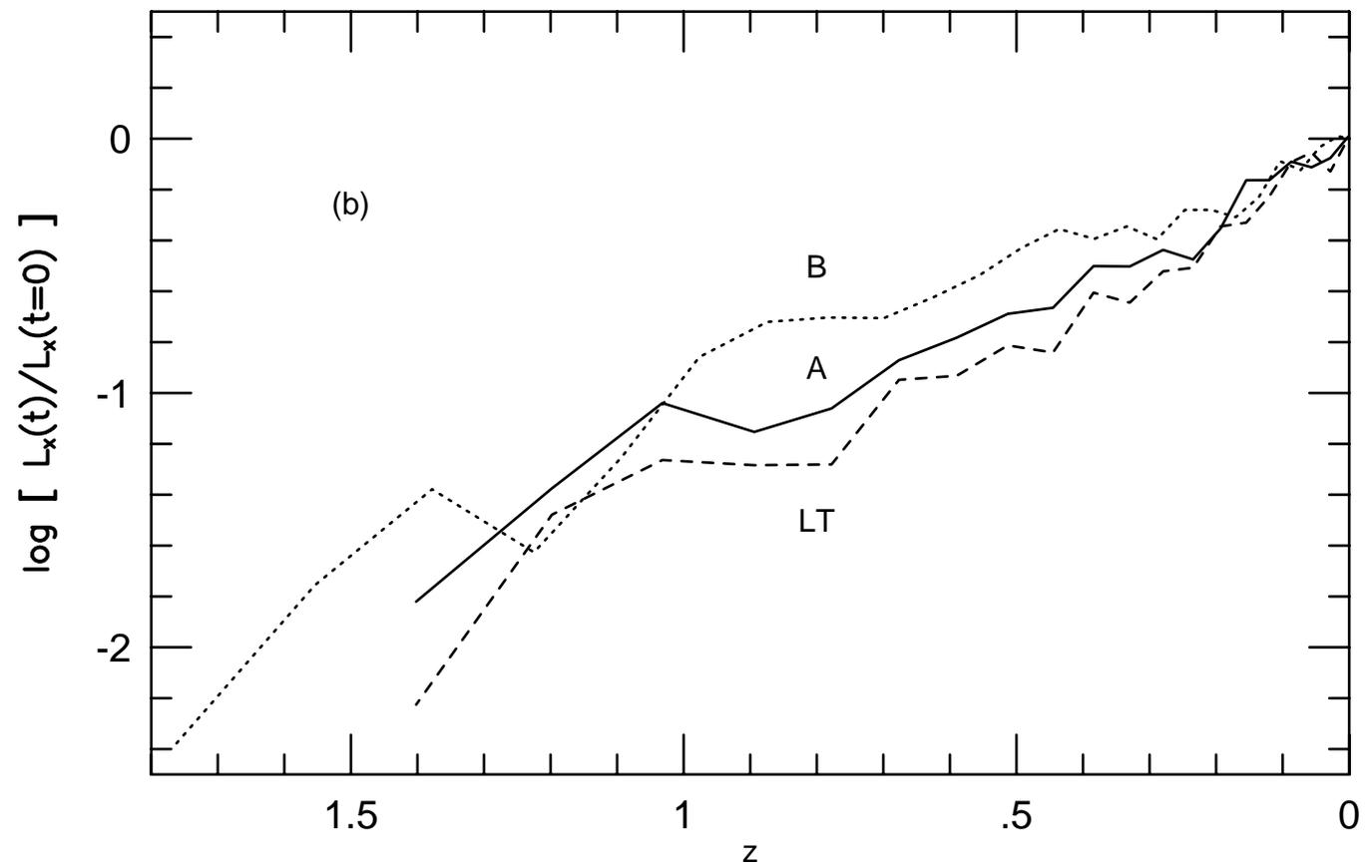
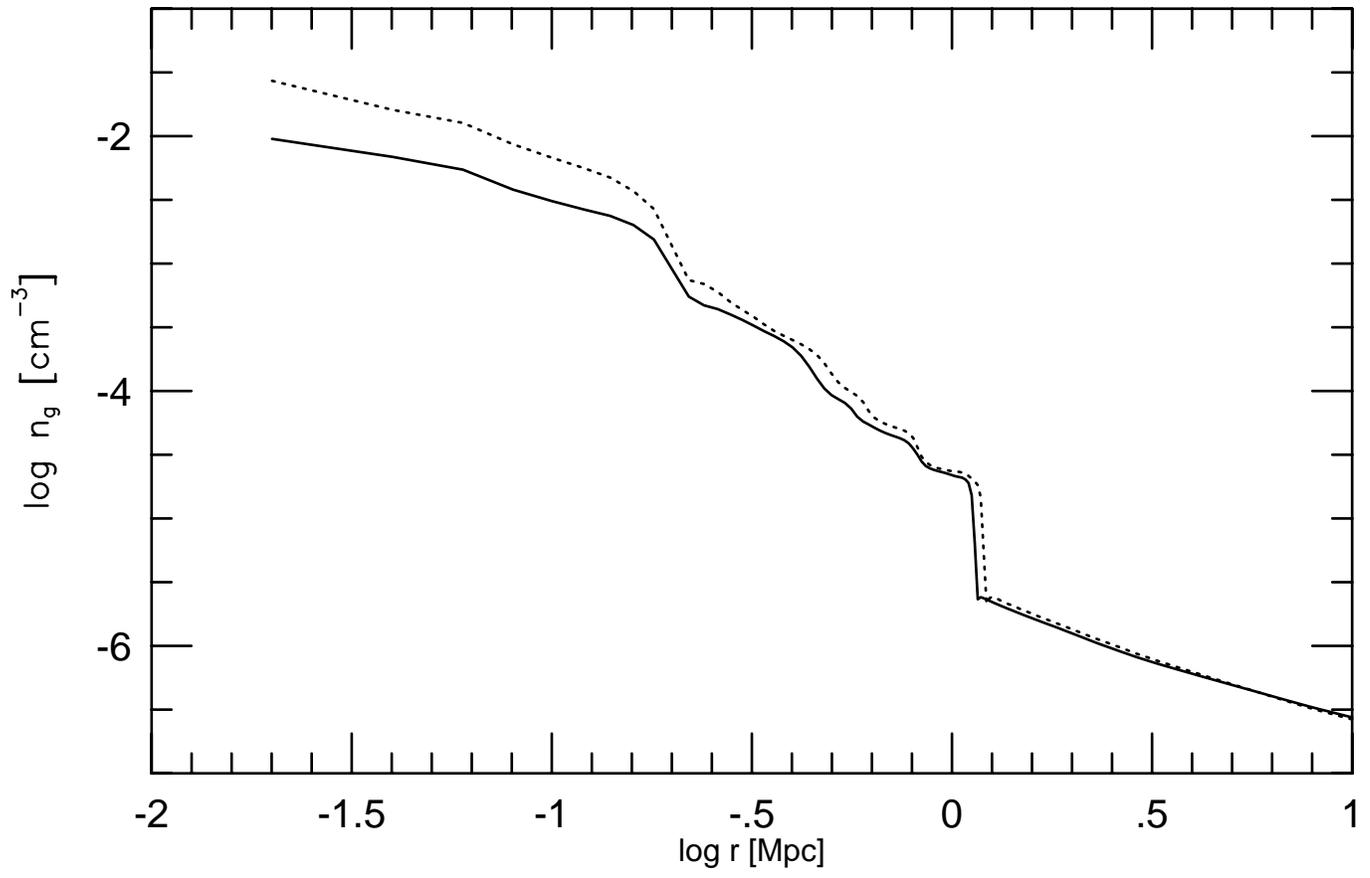
\end{document}